\documentclass[prd,nofootinbib,twocolumn,superscriptaddress,preprintnumbers,balancelastpage,nofootinbib]{revtex4-1}

\usepackage{graphicx}  
\usepackage{dcolumn}   
\usepackage{bm}        
\usepackage{amssymb, amsmath}
\usepackage{slashed}   

\usepackage[usenames,dvipsnames,svgnames,table]{xcolor}

\usepackage[colorlinks=true,linkcolor=Blue,citecolor=Blue,urlcolor=Blue]{hyperref}

\usepackage[utf8]{inputenc}

\def\eps{\epsilon}

\newcommand{\beq}{\begin{equation}}
\newcommand{\eeq}{\end{equation}}
\newcommand{\beqa}{\begin{eqnarray}}
\newcommand{\eeqa}{\end{eqnarray}}

\def\lsim{\mathrel{\rlap{\lower4pt\hbox{\hskip1pt$\sim$}}
    \raise1pt\hbox{$<$}}}         
\def\gsim{\mathrel{\rlap{\lower4pt\hbox{\hskip1pt$\sim$}}
    \raise1pt\hbox{$>$}}}         

\newcommand{\GeV}{{\rm GeV}}
\newcommand{\MeV}{{\rm MeV}}
\newcommand{\TeV}{{\rm TeV}}

\newcommand{\cL}{\mathcal{L}}
\newcommand{\cO}{\mathcal{O}}

\newcommand{\SM}{{\rm SM}}
\newcommand{\BR}{\mathcal{B}}

\newcommand{\Kbar}{\overline{K}^0}
\newcommand{\K}{K^0}

\newcommand{\KOTO}{{\rm KOTO}}
\newcommand{\NA}{{\rm NA62}}
\newcommand{\ECAL}{{\rm ECAL}}

\begin{document}
\widetext

\preprint{KEK--TH--2157, CERN-TH-2019-151}

\title{New physics implications of recent search for  $\boldsymbol{K_L \to \pi^0 \nu\bar{\nu}}$ at KOTO} 

\author{Teppei Kitahara}
\affiliation{Physics Department, Technion---Israel Institute of Technology, Haifa 3200003, Israel}
\affiliation{Institute for Advanced Research \& Kobayashi-Maskawa Institute for the Origin of Particles and the Universe, Nagoya University, Nagoya 464--8602, Japan}

\author{Takemichi Okui}
\affiliation{Department of Physics, Florida State University, Tallahassee, FL 32306, USA}
\affiliation{High Energy Accelerator Research Organization (KEK), Tsukuba, Japan}

\author{Gilad Perez}
\affiliation{Department of Particle Physics and Astrophysics, Weizmann Institute of Science, Rehovot 7610001, Israel}

\author{Yotam Soreq}
\affiliation{Physics Department, Technion---Israel Institute of Technology, Haifa 3200003, Israel}
\affiliation{Theoretical Physics Department, CERN, CH-1211 Geneva 23, Switzerland}

\author{Kohsaku Tobioka}
\affiliation{Department of Physics, Florida State University, Tallahassee, FL 32306, USA}
\affiliation{High Energy Accelerator Research Organization (KEK), Tsukuba, Japan}

\begin{abstract}
\noindent
The KOTO experiment recently reported four candidate events in the signal region of $K_L\to \pi^0 \nu\bar\nu$ search, where the standard model only expects $0.10\pm 0.02$ events. 
If confirmed, this requires physics beyond the standard model to enhance the signal. 
We examine various new physics interpretations of the result including these: 
(1)~heavy new physics boosting the standard model signal, 
(2)~reinterpretation of ``$\nu\bar{\nu}$'' as a new light long-lived particle, or 
(3)~reinterpretation of the whole signal as the production of a new light long-lived particle at the fixed target. 
We study the above explanations in the context of a generalized new physics Grossman-Nir bound coming from the  $K^+ \to \pi^+\nu\bar{\nu}$ decay, bounded by data from the E949 and the NA62 experiments.
\end{abstract}


\maketitle

\section{Introduction}
\label{sec:intro}
 
Despite being one of the greatest successes of theoretical physics, it is clear that the standard model (SM) of particle physics is not a complete description of nature as evidenced by, for example, its lack of a dark matter candidate and a mechanism to produce more matter than antimatter as observed in the universe. Theoretically, the SM suffers from extremely small, unexplained numbers such as the smallness of the electroweak scale compared to the Planck scale ($\sim 10^{-32}$) and the CP-violating vacuum angle associated with the strong nuclear forces ($\lsim 10^{-10}$). One of the best ways to search for new physics (NP) beyond the SM is to look for events that are predicted to be extremely rare in the SM by a theoretically clean calculation. An observation of just a few such events could then constitute a robust evidence of NP\@. A good analogy is the discovery of the positron by Anderson in 1932, for which one event was enough as the expectation from the then ``standard model'' was zero. From this perspective, rare decays of $K$ mesons via a flavor changing neutral current and/or a CP violation (CPV) provide ideal probes of NP as they are highly suppressed in the SM and are theoretically clean~\cite{Littenberg:1989ix}.

Two golden channels are the $K_L\to\pi^0\nu\bar\nu$ and $K^+\to\pi^+\nu\bar\nu$ decay processes.
Within the SM, these are suppressed by a loop factor, the GIM mechanism~\cite{Glashow:1970gm}, and the CKM elements, and predicted to have branching ratios smaller than $10^{-10}$~\cite{Buras:2006gb, Brod:2010hi, Buras:2015qea}.
These processes are being currently probed by the KOTO experiment at J-PARC and the NA62 experiment at CERN, 
both aim to reach the corresponding SM sensitivity. 
Recently,  the KOTO experiment gave a status report for $K_L\to\pi^0\nu\bar\nu$ search~\cite{KOTOslides},
and the NA62 experiment announced new preliminary result for 
$K^+\to\pi^+\nu\bar\nu$ search~\cite{NA62slides}. 

\begin{figure}[t]
\vspace{-0.1cm}
\includegraphics[width=0.4\textwidth]{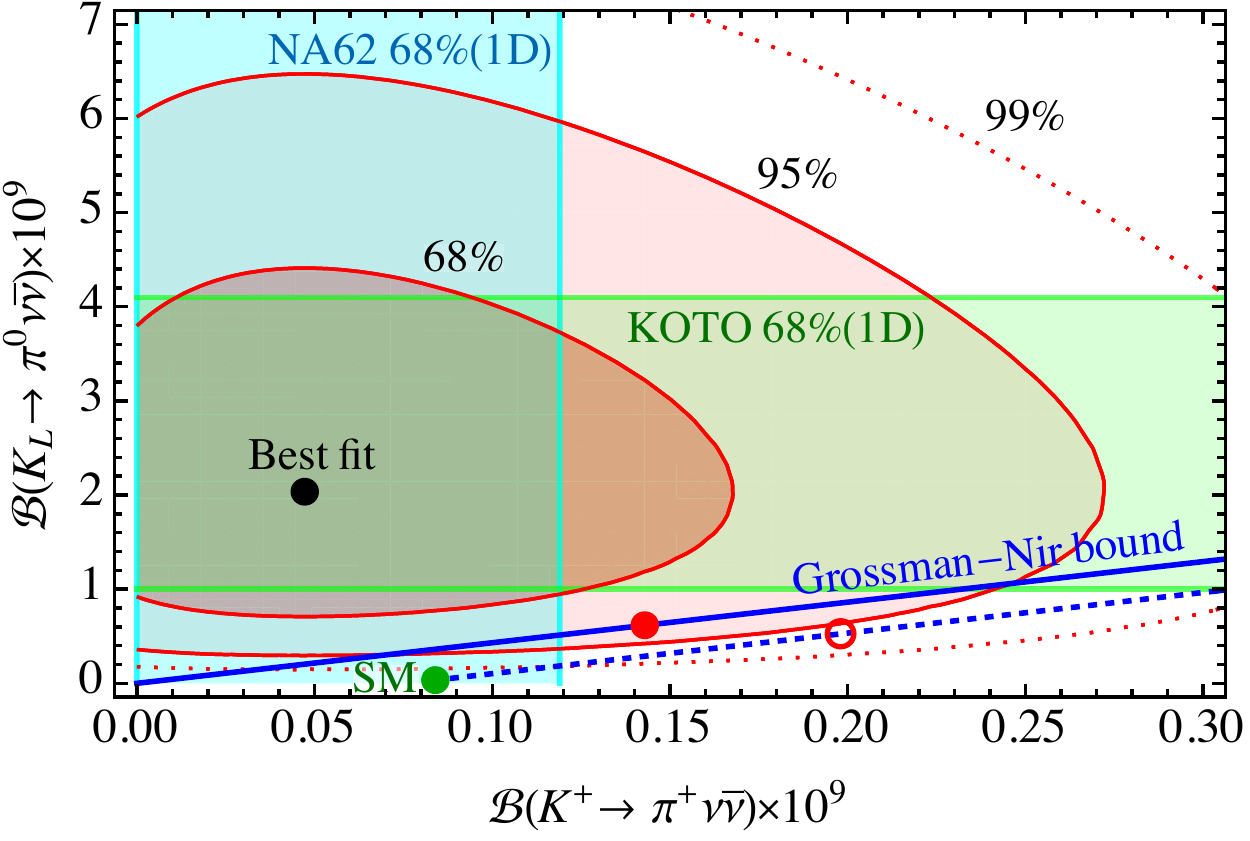}
 \vspace{-0.2cm}
\caption{
The recent result of KOTO events~\cite{KOTOslides}\,(NA62 result~\cite{NA62slides}), Eq.~\eqref{eq:KLKOTO}\,[Eq.~\eqref{eq:BKPNA62}], 
is represented by the green\,(blue) band.
The red ellipses show our simultaneous fits to both.
The GN bound with\,(without) interference with the SM
is shown by the solid\,(dashed) blue line.
The red dots are the best fit points on those lines.
Only statistical uncertainties are taken into account.
}
 \vspace{-0.2cm}
\label{fig:GNKOTO}
\end{figure} 

Strikingly, the KOTO experiment presented data on four
candidate events in the signal region of the
$K_L\to\pi^0\nu\bar\nu$ search,
where the SM expectation is a mere $0.10\pm0.02$ events~\cite{KOTOslides} ($0.05\pm 0.01$  signal and $0.05\pm 0.02$  background). 
While one of the events is suspected as a background from an upstream activity,
the remaining three events are quite distinct from presently known backgrounds. 
In this Letter, we assume that these three events are signals and explore implications, although taking four events as signal would not essentially affect our NP interpretations. 

If the photons and missing energy in the signals are interpreted as $\pi^0 \nu \bar{\nu}$,
the KOTO single event sensitivity, $6.9\times10^{-10}$ \cite{KOTOslides}, implies (for two-sided limits)
\begin{align}
	\label{eq:KLKOTO}
	\BR(K_L \to \pi^0 \nu\bar\nu )_\KOTO  = 2.1^{+2.0 \,(+4.1)}_{-1.1\,(-1.7)} \times 10^{-9}\, ,
\end{align}
at the 68\,(95)\,\%~confidence level~(CL), statistical uncertainties included. 
The central value is about two orders of magnitude larger than the SM prediction, $\BR(K_L\to \pi^0 \nu\bar\nu)_\SM=(3.4\pm 0.6)\times 10^{-11}$~\cite{Buras:2006gb, Brod:2010hi, Buras:2015qea}, 
which corresponds to $p$ value at the $10^{-4}$ level for the SM and background expectations.
On the other hand, for the upper bound on the $K^+ \to\pi^+\nu\bar\nu$ decay rate, the E949 experiment obtained $\BR(K^+ \to \pi^+ \nu\bar\nu) < 3.35 \times 10^{-10}$ at 90\,\%~CL~\cite{Artamonov:2008qb, Artamonov:2009sz}, while the recent preliminary update~\cite{NA62slides} by the NA62 experiment is
\begin{align}
	\label{eq:BKPNA62}
	\BR(K^+ \to \pi^+ \nu\bar\nu)_\NA =0.47^{+0.72}_{-0.47}  (<2.44) \times 10^{-10}\,,
\end{align}
at the 68\,(95)\,\% CL for two-sided (one-sided) limit,
consistent the SM prediction of $\BR(K^+\to \pi^+ \nu\bar\nu)=(8.4\pm 1.0 )\times 10^{-11}$~\cite{Buras:2006gb, Brod:2010hi, Buras:2015qea}.
In Fig.~\ref{fig:GNKOTO}, we summarize the KOTO events and NA62 result (green and blue bands, respectively) and the SM prediction (green dot), and also show our fit to these (red ellipses), 
where in the plot the systematic uncertainties in the backgrounds and the SM theoretical predictions are neglected  as the statistical ones dominate.

We will examine three possibilities to explain the observed events.   
First, we enhance the $K_L \to \pi^0 \nu\bar\nu$ rate by heavy NP.
Such heavy NP can be captured by effective operators, which
we will examine in Sec.~\ref{sec:HeavyNP}.
Second, we interpret  the ``$\nu \bar{\nu}$'' in Eq.~\eqref{eq:KLKOTO} as a new light invisible particle $X$.
We will analyze this scenario in Sec.~\ref{sec:LightNP}.
Interestingly, we will find that the compatibility of the KOTO events and NA62 result require that the $X$ should be a long-lived unstable particle, preferably a scalar, decaying to, {\it e.g.}, two photons.
This may be related to possible solutions to deep problems of the SM, such as the strong CP problem~\cite{Peccei:1977ur,Peccei:1977hh,Weinberg:1977ma,Wilczek:1977pj} or hierarchy problem~\cite{Graham:2015cka,Frugiuele:2018coc,Flacke:2016szy}. 
The last scenario is that  the signals actually have nothing to do with neither $\pi^0$ or $\nu \bar{\nu}$ or not even $K_L$ but are simply due to the production of a new light particle at the fixed target. 
The new particle subsequently decays to two photons after a long flight, where the flight path would generically be off axis and hence appear as ``$\nu \bar{\nu}$.''
While an accurate study of this scenario is challenging as it requires detailed account of the experimental setups, 
we will perform some rough estimates in Sec.~\ref{sec:FixedTarget} to show that it is plausible.

Although the required NP enhancement of the $K_L \to \pi^0 \nu\bar\nu$ rate is substantial to account for the central value of Eq.~\eqref{eq:KLKOTO}, 
 most of other measurements do not have the required sensitivity to directly probe such enhancement.
However, under fairly general assumptions, 
the $K_L \to \pi^0 \nu\bar\nu$ rate can be strongly constrained by the $K_+ \to \pi^0 \nu\bar\nu$ rate via the Grossman-Nir~(GN) bound~\cite{Grossman:1997sk}:
\begin{align}
	\label{eq:GNbound}
	\BR(K_L\to\pi^0\nu\bar\nu) 
	\ \leq \
	4.3 \, \BR (K^+\to\pi^+\nu\bar\nu)\,.
\end{align}
The numerical factor comes from the difference in the total decay widths of $K_L$ and  $K^+$, isospin breaking effects, and QED radiative corrections~\cite{Mescia:2007kn, Buras:2015qea}.
In Fig.~\ref{fig:GNKOTO}, the GN bound 
is shown as the solid (dashed) blue line for NP contributions which interfere (does not interfere) with the SM.

Assuming that the interfering NP$+$SM saturates the GN bound and moving along the solid blue line,
we find that the KOTO and NA62 average deviates at $2.1\,\sigma$ at the red dot on the solid blue line in Fig.~\ref{fig:GNKOTO}. If, instead, we consider the non-interfering case, we have
\begin{align}
	&\BR(K_L\to\pi^0\,\textrm{inv.}) 
	=\>	\BR(K_L\to\pi^0\nu \bar\nu)_\SM
	\nonumber\\
	& \quad + 4.3 \!\left[ \BR (K^+\to\pi^+\,\textrm{inv.}) -  \, \BR (K^+\to\pi^+\nu \bar\nu)_\SM  \right],
\end{align}
where $\textrm{inv.} = \nu \bar\nu$~(SM) $+$ invisible final states~(NP).
In this case, we obtain $2.6\,\sigma$ tension at the red dot on the dashed blue line in Fig.~\ref{fig:GNKOTO}.
A violation of the GN bound by NP contributions is quite difficult (see
Sec.~\ref{sec:disc} for more detail).
In the following, we will not consider the violation of the GN bound.

We shall now discuss in detail the NP scenarios we alluded to above.

\section{Heavy new physics}
\label{sec:HeavyNP}

\begin{figure*}[t!]
\hspace{-0.5cm}
\includegraphics[width=0.46\textwidth]{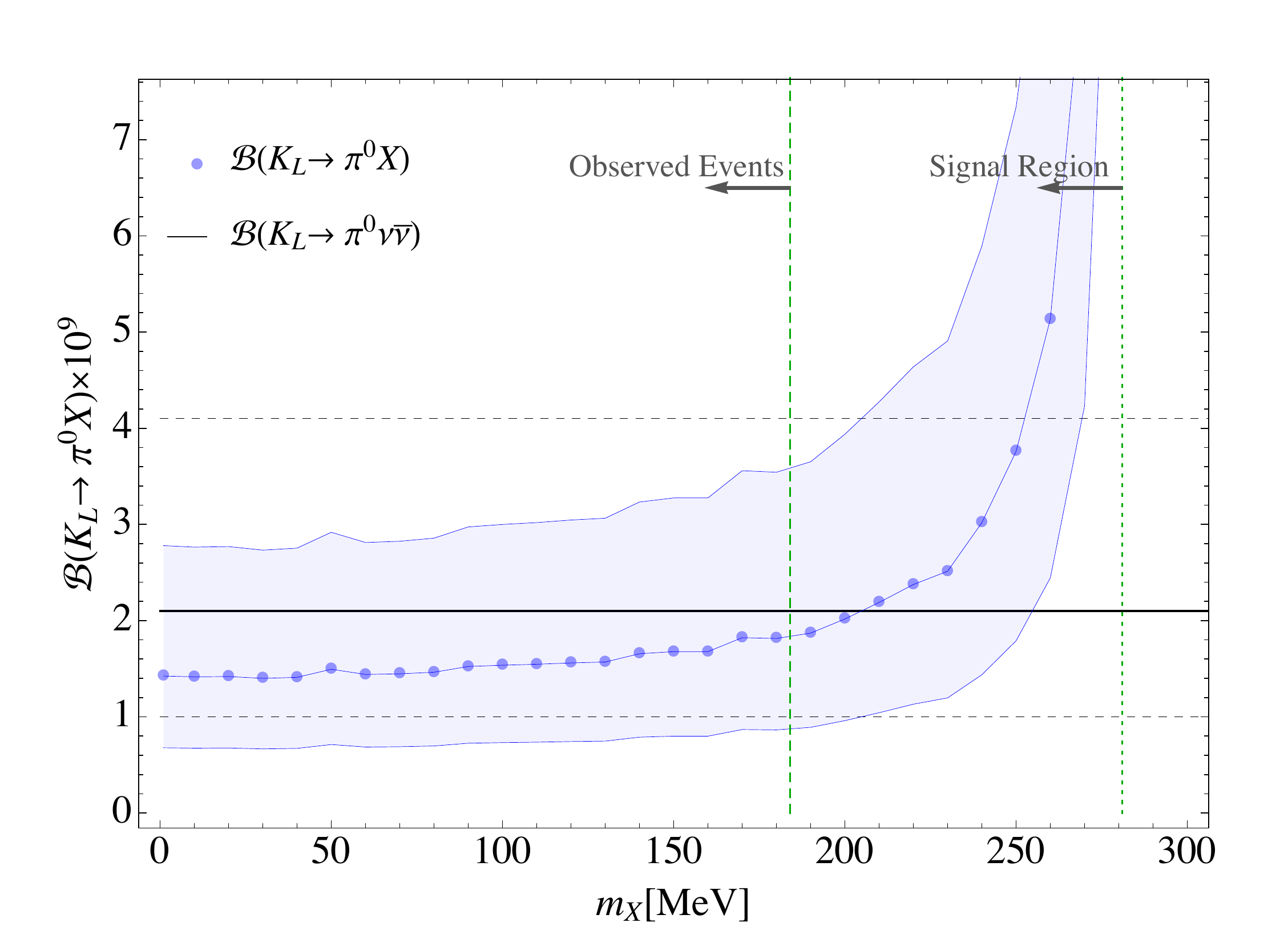}~
\includegraphics[width=0.46\textwidth]{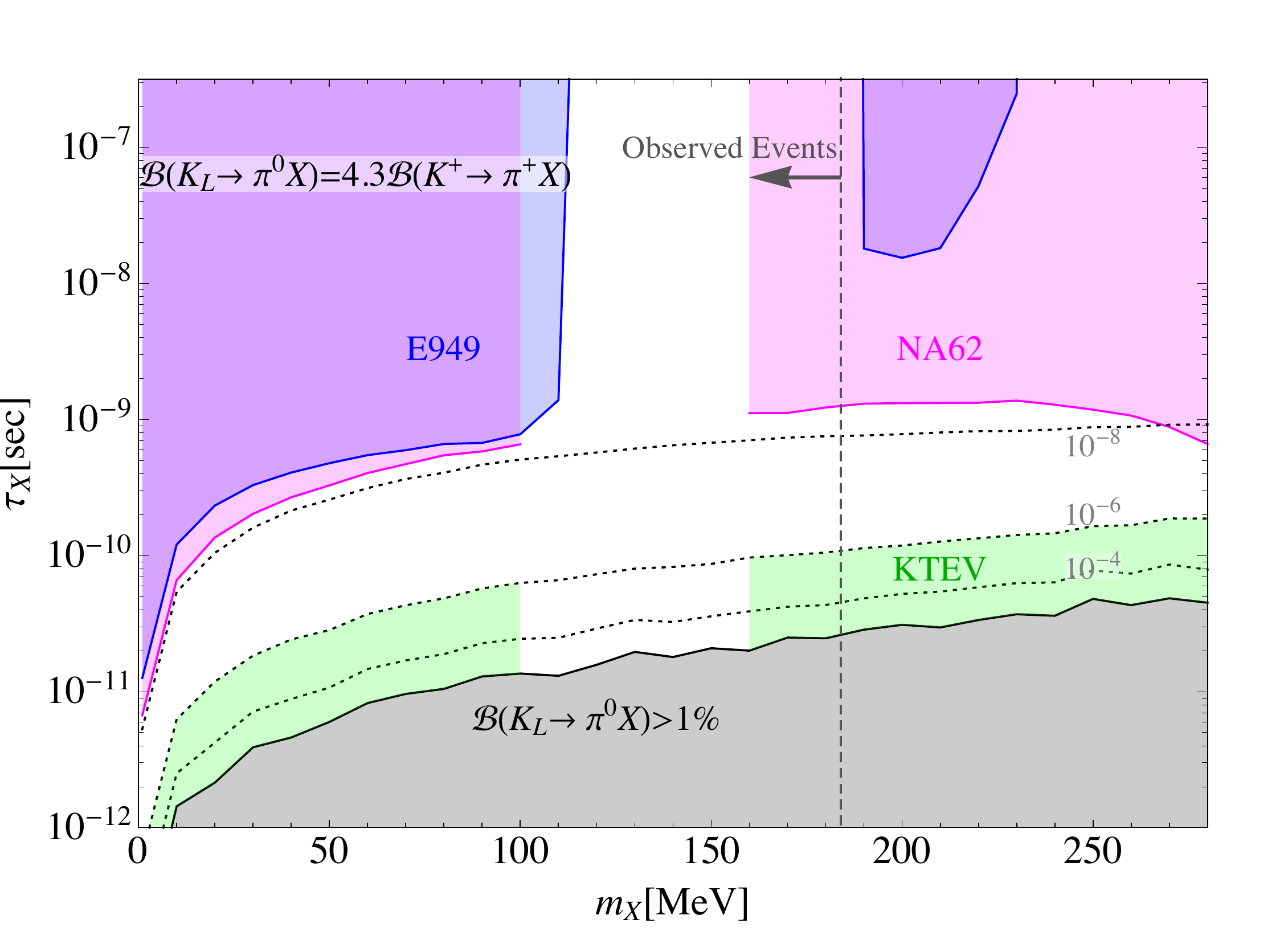}
\caption{
{\bf Left:} branching ratio of $K_L\to \pi^0\nu\bar\nu$ or $\pi^0X$ that can accommodate the KOTO events, see Eqs.~(\ref{eq:KLKOTO}) and~\eqref{eq:KLKOTO2body}. 
The dotted blue\,(solid gray) line  correspond to the central value of $K_L\to \pi^0 X\,(\pi^0\nu\bar\nu)$ interpretation, with blue shaded band\,(dashed horizontal lines) for two-sided 68\% confidence interval.   
An uncertainty of Monte Carlo statistics is less than 10\% thus omitted here.  
The dashed (dotted) vertical line corresponds to
$m_X = 180\,(280)\,\MeV$, and its left-hand side is compatible with the observed events (the signal region).
{\bf Right:} the new particle has finite lifetime considering the GN bound and $K^+\to \pi^+\nu\bar\nu$ search, and the allowed parameter space for two body decay $K_L\to \pi^0X$ in mass and lifetime of $X$ is shown. See~Eq.~\eqref{eq:K2piinv2}. 
The $K^+ \to \pi^+X$ bound is translated to $K_L$ bound assuming a saturation of the GN bound.
The purple\,(blue) shaded region is constrained by NA62~\cite{NA62slides} at 95\,\%~CL\,(E949~\cite{Artamonov:2009sz} at 90\,\%~CL). 
Too short lifetime leads to
$\BR(K_L \to \pi^0 X ) > 1\%$, which is inconsistent with {\it untagged} $K_L$ branching ratio~\cite{Tanabashi:2018oca}.  
The $\BR(K_L\to\pi^0 X)=10^{-4},\, 10^{-6}$ and $10^{-8}$ are indicated on the plot.
The green shaded region is constrained from KTEV search for $K_{L} \to \pi^0 \gamma \gamma$ assuming $\BR(X\to\gamma\gamma)=1$~\cite{Abouzaid:2008xm}.
}
\label{fig:BrpiXKOTO}
\end{figure*} 

First, let us consider heavy NP which contributes to $s \to d \nu \bar\nu$ processes.
Matching the fields involved in the $K_L\to \pi^0\nu\bar\nu$ decay to a gauge invariant dimension-six operator, the effective Lagrangian, with operators that can interfere with the SM contributions, only consists of three operators, 
$\cL_{\rm eff} =  \sum_{i=S,A,D}C_i^{\nu\nu} \cO_i^{\nu\nu} + \textrm{h.c.}$ with $\cO_{S,A}^{\nu\nu} =\left[\bar Q^2\left(\mathbf{1}_2,\sigma^i\right)Q^1\right]_{V-A} \left[\bar L \left(\mathbf{1}_2,\sigma^i\right)L\right]_{V-A}$ and $\cO_{D}^{\nu\nu} = \left(\bar{d}^2d^1\right)_{V+A} \left(\bar L L\right)_{V-A}$ 
where $Q$\,($L$) is a quark\,(lepton) doublet, $d$ is the down-type quark singlet, 
$\mathbf{1}_2$ and $\sigma^i$ are in SU(2$)_L$ weak space, the superscripts 1 and 2 correspond to quark-generation index in the down mass basis and  lepton flavor indices are suppressed for here. 
 For example, these operators can be a low energy description of a flavorful $Z'$ model.

By considering the single complex Wilson coefficient $C_{S,A,D}^{\nu\nu}$ (defined at the $m_Z$ scale), and fitting it to separately  the KOTO events and then both to KOTO and NA62 to minimize the tension between the experiments we find, %
\begin{align}
	\label{eq:WCfit}
	\! \!\!\! \! 
	C_{S,D}^{\nu\nu}   - C_{A}^{\nu\nu} \!\approx \!
	\left\{	
	\begin{array}{ll}
	\mathrm i/ (110 \,{\rm TeV})^2, & \textrm{KOTO} \\ 	
	\mathrm{e}^{ - \mathrm{i} {3\over 4}\pi }/(150 \,{\rm TeV})^2, &\textrm{KOTO}\, \&\,\textrm{NA62}
	\end{array}
	\right. \! ,
\end{align}
where the value on the first line of the above equation corresponds to fitting for the central value of KOTO only, and on the second line we fit  both to the KOTO events and NA62 result,  which corresponds to the red solid dot in Fig.~\ref{fig:GNKOTO}.

Assuming lepton flavor universality, the above operators can be sensitive to CP-violating flavor changing neutral current 
such as $K_L \to \pi^0 \ell^+ \ell^-$ ($\ell = e, \mu$) and $K_S \to \mu^+ \mu^-$, whose branching ratios are experimentally bounded as $\lsim \text{(a few)} \times 10^{-10}$~\cite{AlaviHarati:2003mr, AlaviHarati:2000hs, LHCbslides}.
In light of the fact that $K_L\to\pi^0\nu\bar\nu$ search is the neutrino flavor blind, these upper bounds would be the same order as the predictions of Eq.~\eqref{eq:WCfit}.
If the NP couples only to one neutrino flavor, 
the scale of  Eq.~\eqref{eq:WCfit} will barely change.
In particular, it would be interesting to consider a correlation with the direct CPV in  $K^0 \to \mu^+ \mu^-$~\cite{DAmbrosio:2017klp,Chobanova:2017rkj} which would be probed by the LHCb experiment.
However, these bounds can be avoided if one is  switching on the coupling to third generation leptons only in the definition of $\cO_{S,A,D}^{\nu\nu}\,.$ 
Additional option to avoid these could be found by making the  coefficient of $\cO_{S}^{\nu\nu}$ and $\cO_{A}^{\nu\nu}$   
 to obtain ``custodial symmetry" so that the coupling to the charge lepton bilinear is switched off~\cite{Agashe:2006at}. 
In this case, a potentially interesting effect would appear via charged current in  the decay of $\tau\to K(\pi)\nu$, where there is a $2.8\,\sigma$ tension in the CP asymmetry of $\tau \to K_S \pi^- \nu_{\tau}$~\cite{Grossman:2011zk,Cirigliano:2017tqn}.
However, since flavor changing charged current occurs at tree level in the SM, such $\cO(100)$\,TeV scale sensitivity as in Eq.~\eqref{eq:WCfit} is unlikely. From the same reason correlated transition involves charm decay are also expected to lead to subdominant effects that are hard to observe (see however~\cite{Gedalia:2012pi}).
Furthermore, one can obviously assume non-universal lepton interactions and switch off the couplings to the tau or to other charge leptons.

Finally we comment that one can also account for the above decay by adding operators with right-handed neutrino field $N$ of the form $\bar Q^2 d^1 \bar L N$, $\bar Q^2 \sigma_{\mu\nu} d^1 \bar L\sigma^{\mu\nu} N$ (plus $\bar Q^2 d^1 \leftrightarrow \bar Q^1 d^2$)
and $\left(\bar Q^2 Q^1\right)_{V-A} \left(\bar N N\right)_{V+A}$, where the correlation with charged lepton signal becomes weaker or can be avoided altogether.
As these operators do not interference with the SM, they would result in a stronger tension with the data.
In this case, the best fit point corresponds to the empty red point of Fig.~\ref{fig:GNKOTO}.

\section{Light New Physics}
\label{sec:LightNP}

The observed KOTO events could be explained by a two body decay associated with a new invisible particle, $X$; $K_L \to \pi^0 X$. 
Below we show that
 the new particle cannot be completely invisible but \emph{must decay with a finite lifetime}
of $\mathcal{O}(0.1$--0.01)\,ns, except for $m_X \approx m_{\pi^0}$.

For $m_X < m_{\pi^0}$, 
$K_L\to\pi^0X$ and $K_L\to\pi^0\nu\bar\nu$ have similar kinematic features in the KOTO signal region, thus, 
the required branching ratio to explain the KOTO events is $\sim 10^{-9}$, 
similar to Eq.~\eqref{eq:KLKOTO}. 
For $m_X > m_{\pi^0}$, 
the signal efficiency will be reduced, thus, more events are required.
The reconstructed pion transverse momentum in the signal region must be in the range $130~\MeV < p_T^{\pi^0}<250~\MeV$, while 
the transverse momentum from the $K_L \to \pi^0 X$ decay is limited by the phase space as 
$p_{T,{\rm max}}^{\pi^0}=\sqrt{ \lambda(m_X^2 , m_{K_L}^2, m^2_{\pi^0} )} / 2m_{K_L}$ with $\lambda (a,b,c) = a^2 +b^2 + c^2 - 2 (ab + bc + ca)$. 
Ignoring detector effects, the signal of the two-body decay with $m_X>280~\MeV$ $(p_{T,{\rm max}}^{\pi^0}<130~\MeV)$ does not overlap with the KOTO signal region. 
It is notable that all of the three KOTO events in question have $p_T^{\pi^0}\gtrsim 190\,\MeV$,  which may indicate that $m_X\lesssim180~\MeV$ is favored.

To take into account the efficiency difference from $\pi^0\nu\bar\nu$, we correct 
Eq.~\eqref{eq:KLKOTO} by the ratio of efficiencies estimated by our simulation as 
\begin{align}
	\label{eq:KLKOTO2body}
	\frac{\BR(K_L \to \pi^0 X)_\KOTO}{\BR(K_L \to \pi^0 \nu\bar\nu )_\KOTO}=
	\frac{\eps_{\pi^0 \nu\bar\nu}}{\eps_{\pi^0 X}(m_X)}\,,
\end{align}
where $\eps$ is the efficiency of kinematic cuts of an earlier KOTO analysis \cite{Ahn:2018mvc} and new signal region of reconstructed momentum and decay vertex \cite{KOTOslides}. 
The result is shown in the left panel of Fig.~\ref{fig:BrpiXKOTO}. The simulation setup and validation are  presented in Appendix~\ref{simulation}.

Similarly to the $K_L\to\pi^0\nu\bar\nu$ case, the rare decay of $K^+$ search will constrain this scenario. 
This is because, even in this case, a generalized version of the GN  bound still holds~\cite{Leutwyler:1989xj}, 
\begin{align}
	\label{eq:KpiX}
	\BR(K_L\to\pi^0X)\ \lesssim\ 4.3\, \BR(K^+\to\pi^+X)\,.
\end{align}
The upper bound on two body decay $\BR(K^+\to\pi^+X) $ is $\cO(10^{-10}\textrm{--}10^{-11})$~\cite{Artamonov:2009sz}, which is generally stronger than  that on $\BR(K^+\to\pi^+\nu\bar\nu)$ except for near the neutral pion mass $|m_X-m_{\pi^0}|\lesssim 25\,\MeV$  and above two pion mass threshold $m_X\gtrsim 2m_{\pi^0}$, because the search is suffered from $K^+\to \pi^+\pi^0(\gamma),\, \pi^+ +2\pi$ backgrounds \cite{Artamonov:2009sz, NA62slides, CortinaGil:2018fkc, Fuyuto:2014cya}.  
For example, for $m_X=(0)\,100\,\MeV$, the expected number of events in KOTO is bound to be smaller than $0.7\,(0.3)$ at $90\,\%$~CL. 

This situation is changed when the invisible particle $X$ is unstable and can decay into the visible particles such as photons. 
Once $X$ decays, say, to photons, the events are vetoed or go to different search categories where the bound on branching ratio is significantly weaker due to large SM contributions of $K_L(K^+)\to \pi^0(\pi^+)+ \pi^0$ or $\pi^0(\pi^+)+\gamma\gamma$ with Refs.~\cite{Abouzaid:2008xm, Artamonov:2005ru, Ceccucci:2014oza}. 

The dependence of the efficiency on $X$ lifetime of $K_L\to\pi^0X$ is different than that of $K^+\to\pi^+X$ because the boost factors, $p/m_X$ and the effective detector size, $L$ of NA62 or E949, which are different than those of KOTO. 
Effective branching ratios are  
\begin{align}
	\label{eq:K2piinv}
	\BR(K\!\to\!\pi\!X;{\rm detector})
=	\BR(K\to\pi X) e^{ -\frac{L}{p}\frac{m_X}{c\tau_X} }  \,,
\end{align}
which are measured by experiments. 
Through the GN bound, Eq.~\eqref{eq:KpiX}, the bound on the lifetime is obtained by taking a ratio,  
\begin{align}
	\label{eq:K2piinv2}
	\frac{\BR(K^+ \to \pi^+X)_\NA^{\rm 95\%CL}}{{\cal B}(K_L \to \pi^0 X)_\KOTO} >
	\frac{\BR(K^+\to\pi^+ X{\rm ; \NA})}{{\cal B}(K_L \to \pi^0 X{\rm ;\KOTO})} \nonumber \\
	\geq \frac{1}{4.3}\exp\left[ -\frac{m_X}{c\tau_X}\left(\frac{L_\NA}{p_\NA} - \frac{L_\KOTO}{p_\KOTO} \right) \right]\,,
\end{align}
where we use the central value of Eq.~\eqref{eq:KLKOTO2body} and the bound ${{\cal B}(K^+ \to \pi^+X)_{\rm NA62}^{\rm 95\%CL}=1.6 \times 10^{-10}}$   which is the NA62 bound [Eq.~\eqref{eq:BKPNA62}] subtracting  non-interfering SM contribution. 
The exponential factor is calculated by simulation for KOTO using the selected event samples in the signal region.
 To a good approximation,
  one can use $L\simeq 3\,$m and  $E_X\simeq 1.5\,\GeV$ and  
for NA62, we take $E_X=37\,\GeV$ and $L=150\,$m.  
Because effective detector size of KOTO is smaller than that of NA62, $L_\NA/p_\NA > L_\KOTO/p_\KOTO$, the bound of NA62 can be evaded for some shorter lifetime. If the lifetime is too short, roughly less than 0.01\,ns,  the branching ration of $K_L\to \pi^0 X$ has to exceed $1\%$, which is constrained by sum of the other decay channels of $K_L$. 
For E949, we can write the analogous formula, and there the $K^+$s are at rest, thus $p_X$ is calculated  and $L=1.5\,$m. Because the $p_X$ is much smaller, the effective detector size ${L_{E949}}/{p_{E949}}$ is much larger  than that of KOTO and NA62 especially for higher mass, making NA62 more sensitive to this scenario.  The experimental bound of E949 uses Fig.~18 of Ref.~\cite{Artamonov:2009sz}. 
The results are shown in the right panel of Fig.~\ref{fig:BrpiXKOTO}. 

Assuming the GN bound is saturated ${\cal B}(K_L\to\pi^0X)=4.3\,{\cal B}(K^+\to\pi^+X)$, we found that parameter space of the lifetime  $\cO$(0.1-0.01)\,ns is compatible with both KOTO and NA62\,(E949). 
Using  visible decay channels such as ${\cal B}(K_L\to \pi^0X,X\to 2\gamma)$ \cite{GORIPEREZTOBIOKA},  if one will find the favored lifetime is inside the parameter space excluded by $K^+\to\pi^+X$, it indicates the violation of  the GN bound. For a constraint from the visible channel, KTEV $K_L\to \pi^0\gamma\gamma$ will exclude $\BR(K_L \to \pi^0 X)\gtrsim10^{-6}$ if $X$ decays dominantly to two photons \cite{Abouzaid:2008xm}. 

Let us comment on possible underlying models of $X$.
Arguably the simplest possibility is a Higgs portal which induces $K_L\to \pi^0 X$ decay, but the dominant decay of $X$ is into $e^+e^-$ which is tightly constrained by KTEV search, ${\cal B}(K_L \to \pi^0 e^+ e^-)<2.8 \times 10^{-10}$ at 90\%\,CL \cite{AlaviHarati:2003mr}. 
One can avoid this bound easily if the $X$ is some kind of leptophobic and/or photophilic scalar. 
For example, if there are two (or more) Higgs doublets, one Higgs is responsible to the masses of third generation and quarks, another one is responsible to the masses of light leptons, and $X$ mixes with just the former Higgs.

\section{New Particle Production at Fixed Target}
\label{sec:FixedTarget}

An alternative scenario that could accommodate the KOTO 
events is that  the events are not due to an enhanced $K_L \to \pi^0 + \text{(inv.)}$ rate but just a disguise of a new light particle, $\phi$, produced at the fixed target and decaying inside the vacuum chamber to a photon pair.
At KOTO, the initial 30\,GeV proton beam hits the fixed gold~(Au) target at an angle of $16^\circ$ with respect to the beam line connecting the target and the electromagnetic calorimeter~(ECAL). 
Unlike the $K_L$, which would travel straight along the beam line toward the vacuum chamber, the new particle will not fly parallel to the beam line so it will enter the chamber away from the axis with an angle.  
We further assume that the $\phi$ lifetime is such that it typically decays inside the vacuum chamber to two photons.
Moreover, in the $K_L \to \pi^0\nu\bar{\nu}$ search, KOTO does not reconstruct the $\pi^0$ mass but instead assumes that the photon pair detected on the ECAL has an invariant mass of $\pi^0$ and that the pair comes from a vertex on the beam line, for these two assumptions would completely determine the location of a $K_L$ decay to $\pi^0 \nu \bar{\nu}$.

Therefore, we see that the $\phi$'s in-flight decay to $2\gamma$ will indeed disguise as an $\pi^0 +$(invisible) event. 
The kinematics is similar to CV-$\eta$ background, a decay of $\eta\to 2\gamma$ in the off-axis region can have a reconstructed vertex inside the signal region.  
On the other hand, at NA62, which triggers events by charged particles and is designed to veto huge $\pi^0$ background, such $\phi$ decays are simply rejected.
As a concrete example, we consider that $\phi=a$ is an axion like particle~(ALP) with the following effective interactions 
\begin{align}
	\cL_{\rm int}
=	\frac{\alpha_s }{8\pi f_g} a G_{\mu\nu}^a\tilde{G}_{a\mu\nu}+\frac{\alpha_{\rm EM}  }{8\pi f_\gamma}aF_{\mu\nu}\tilde{F}^{\mu\nu} \,  , 
\end{align}
where $F_{\mu\nu}\,(G_{\mu\nu}^a)$ is the photon\,(gluon) field strength and $\tilde{F}_{\mu\nu}(\tilde{G}^a_{\mu\nu})$ is the field strength dual. 
$f_g$ and $f_{\gamma}$ are the decay constants. 
For recent relevant review see Ref.~\cite{Beacham:2019nyx}. 
We consider the case of $m_a<3\,m_{\pi}$ to avoid hadronic decay channels. 
The ALP lifetime is controlled by the photon coupling and IR contribution from the gluon coupling which are in the same order if $f_g\sim f_\gamma$. 
In $p$--Au collisions, ALP can be produced by different mechanisms:
non-perturbative 
production, deep-inelastic scattering, {\it e.g.}, $gg\to g a$, coherent proton-nucleon production, and bremsstrahlung. 
Here we  consider only non-perturbative 
production inferred from the measured $K_L$ flux at KOTO.

The number of decays inside the KOTO detector is
\begin{align}
	N_a
\!=\!	\int \! dp \! \int_{\Delta\Phi_{\rm det}} \!\!\!\!\!\!\! d\Phi 
	\frac{d^2 N_{p{\rm Au}\to a}}{d p~d\Phi} 
	\left[ e^{-\frac{d-L}{p}\frac{m_X}{ c\tau}}-e^{-\frac{d}{p}\frac{m_X}{ c\tau}} \right] \, ,
\end{align}
where $\Delta\Phi_{\rm det}\approx \pi r^2_\ECAL/{4\pi (d+L)^2}\sim 10^{-4}$ is the angular coverage of the detector in the lab frame, with $r_\ECAL\simeq1\,$m is the ECAL radius, $d\simeq 27\,$m is the distance to the ECAL and $L\simeq3\,$m is the distance from photon veto detector to ECAL.
The number of detected events is $N_a \times  A\epsilon $ where $A\epsilon$ is acceptance times reconstruction efficiency for ALPs in the signal region of $K_L\to \pi^0 \nu\bar\nu$. 
Estimating reconstruction efficiency is not trivial because the topology of this signal differs from that of $K_L\to\pi^0\nu\bar\nu$ assumed by KOTO. The precise estimation is left for future work. 

If the ALP mass is below the QCD confinement scale, the gluon interaction induces the ALP-pion mixing with a mixing angle of  $\sin(\alpha_{a\pi})\sim \frac{m_d-m_u}{m_d+m_u}\frac{f_\pi}{2f_g}\frac{m^2_a}{m^2_a-m^2_\pi}$~\cite{Georgi:1986df} (also see, {\it e.g.},~Refs.~\cite{Bauer:2017ris,Aloni:2018vki}).  
Thus, the ALP production rate can be estimated as $dN_{p{\rm Au}\to a}\approx\sin^2(\alpha_{a\pi}) d N_{p{\rm Au}\to \pi^0}$. 
Although there is no data of $N_{p{\rm Au}\to \pi^0}$ for our purpose, we can estimate it the using measured $K_L$ based on a generic expectation  $N_{p{\rm Au}\to \pi^0} > N_{p{\rm Au}\to K_L}$. 
To obtain the differential distribution at the point of $p$--Au collision, we unfold the measured distribution at the beam exit by 
\begin{align}
	\label{eq:KLdist}
	&\int_{\Delta\Phi_{\rm det}} \!\!\!  d\Phi 
	\frac{d^2 N_{p{\rm Au}\to K_L}}{d p~d\Phi} \nonumber
	\\
	&\approx \frac{\Delta\Phi_{\rm det}}{\Delta\Phi_{\rm beam}}
	\int_{\Delta\Phi_{\rm beam}} \!\!\!\!\! d\Phi 
	\frac{d^2 N^{\rm exit}_{p{\rm Au}\to K_L}}{d p~d\Phi} e^{\frac{d_{\rm exit}}{p}\frac{m_{K_L}}{ c\tau_{K_L}}} \, ,
\end{align}
where the angular dependence is neglected within the small region,  $d_{\rm exit}=20\,$m, and the angular coverage of the beam hole $\Delta\Phi_{\rm beam}$ is $(8.5\,{\rm cm})^2/4\pi d_{\rm exit}^2$.  
The momentum distribution is given in Ref.~\cite{Masuda:2015eta} and the normalization of $N^{\rm exit}_{p{\rm Au}\to K_L}$ is fixed $7.1\times 10^{12}$~\cite{KOTOslides}. 
Note that there is an enhancement of $\frac{\Delta\Phi_{\rm det}}{\Delta\Phi_{\rm beam}}\sim 200$  in $N_{K_L}$ yield for our purpose.
Base on the above estimation,  we get $N_a A\epsilon\sim\cO(10^3\text{--}10^5) (f_g/1~\TeV)^{-2}(N_{\pi^0}/N_{K_L})A\epsilon$ at $m_a=200~\MeV$ leading  to $f_g \sim $0.1\textrm{--}1\,TeV for $A\eps=10^{-4}$, $N_{\pi^0}=N_{K_L}$ and $\cO(1)$ events.  
The lifetime can vary from 0.1\,ns to 1\,$\mu$s. 

This is a proof of concept that the scenario can explain the KOTO events, however, a more careful study including a comparison to other existing constrains, see, {\it e.g.},~\cite{Beacham:2019nyx,Aloni:2019ruo,Ebadi:2019gij,Aloni:2018vki,Altmannshofer:2019yji} is required after exploring the value of $A\eps$. 
This type of scenario could be further constrained by beam-dump experiments, in particular the ones using proton beam such as CHARM~\cite{Bergsma:1985qz} and NuCal~\cite{Blumlein:1990ay}. These would constrain parameter space with lifetime above 1\,ns, and the detailed analysis will be presented in the future work.

\section{Discussion}
\label{sec:disc}
%
As alluded to in the Introduction,  we discuss the GN bound  in the presence of NP effects.
The GN bound relies on the following assumptions~\cite{Grossman:1997sk}.  
First, isospin symmetry, which relates the decay amplitudes of $K^\pm$ to the ones of $\K$ and $\Kbar$.  
Second, the ratio of the $\K$ and $\Kbar$ decay amplitudes to the corresponding sum of final states  is close to unity, where if the final state is CP eigenstate it means no CPV in the decay.
 For the $\pi\nu\bar\nu$ final state, within the SM, it is expected to be an excellent approximation.

The above assumptions are not easy to be violated even by NP.
For example, within effective field theory models, isospin violation to the above processes receives leading contribution at dimension-six operators as we need to mediate transition between an isospin doublet to a triplet, 
which involves four-quark operators.
If we further would like to couple it to an ALP  
to boost the $K_L\to \pi^0 a$ rate, 
we arrive at dimension-eight operator  $\partial_\mu a\, (\bar s \gamma^\mu d) (\bar uu-\bar d d)$.
However, axion models are subject to stringent bounds, from flavor, beam dump experiments and astrophysics cooling bounds, because they induce dimension-five operators.
One may hope to violate the GN bound by adding large contribution to CPV in decay.
However, 
 it requires both strong and weak phases to be present, and as the final states involve neutrino (or other SM singlets) the strong phases are generically expected to be suppressed.
Finally, as the signature of the above decay is rather inclusive, basically  looking at a pion and a missing energy, it typically involves summing up all the particles in the final states, which washes away effects 
that distinguish between the decay of $\K$ to some exclusive final state and $\Kbar$ to another different state.
While we are not aware of a basic principle within local quantum field theory framework, such as CPT or unitarity conservation, that guarantees that the GN bound is respected,
we are not aware either of an existing proof for its violation.

\section*{Acknowledgments}
We would like to thank Yossi Nir for valuable discussions and for Jure Zupan for comments on the manuscript.
T.K. is supported by the Israel Science Foundation (Grant No.~720/15), by the United-States-Israel Binational Science Foundation (BSF) (Grant No.~2014397),  by the ICORE Program of the Israel Planning and Budgeting Committee (Grant No.~1937/12), and by the Japan Society for the Promotion of Science (JSPS) KAKENHI Grant Number 19K14706.
G.P~is supported by grants from the BSF, ERC, ISF, Minerva Foundation, and the Segre Research Award.
K.T.~and T.O~are supported by the US Department of Energy grant DE-SC0010102.
K.T. is supported by his startup fund at Florida State University~(Project id: 084011- 550-042584).

\appendix

\begin{figure}[t!]
\vspace{0.5cm}
\includegraphics[width=0.48\textwidth]{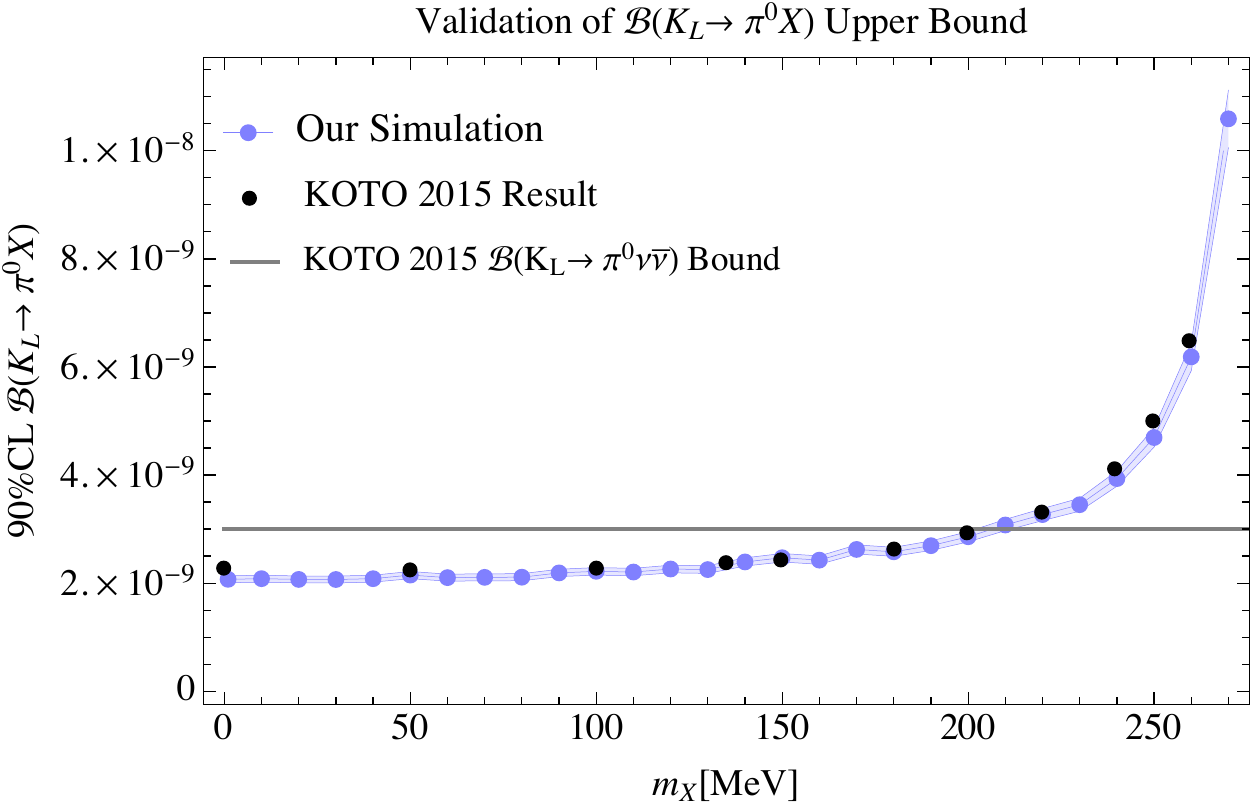}
\caption{Comparison for $K_L\to \pi^0 X$ upper bound for the validation of our simulation. The black dots are from Fig.~4 of~\cite{Ahn:2018mvc}, while the blue dots are derived by Eq.~\eqref{eq:KLKOTO2body} using $\BR(K_L\to\pi^0\nu\bar\nu)^{ }_{\rm KOTO}<3.0\times 10^{-9} $ at 90\% CL (gray horizontal line). The blue band is statistical uncertainty of our MC sample size. 
 }
\label{fig:shape}
\end{figure} 

\section{Validation of our $\pi^0 X$ simulation for KOTO}
\label{simulation}

Here, we would like to describe and validate the simulations we performed to evaluate the efficiency factors in Eqs.~(\ref{eq:KLKOTO2body}) and (\ref{eq:K2piinv}). 
 The KOTO experiment has a simple setup that the $K_L$ beam enters the detector and the two photons from the $\pi^0$ from a $K_L$ decay then hit the ECAL\@. 
This allows a simple simulation to reproduce the signal shape well \cite{GORIPEREZTOBIOKA}.  
Determination of a decay vertex is a challenge for KOTO as all particles involved in $K_L\to \pi^0\nu\bar\nu$ and $\pi^0\to2\gamma$ are charge neutral, and there are no directional information. 
However, assuming that the $K_L$ has decayed on the beam axis and the two photons are from the $\pi^0$  from the $K_L$, one can reconstruct the decay with a very good accuracy \cite{Masuda:2015eta}. 

We first generated initial $K_L$ particles following the measured momentum distribution \cite{Masuda:2015eta}. Then, we let each $K_L$ decay between ECAL position and the entrance of the detector (4.148\,m upstream from the ECAL). 
We found 7.9\% of $K_L$s decays in this region compared to the number of $K_L$s at the beam exit (at the end of the second collimator). 
We treated every two-body decay as isotropic.  For three-body decay $K_L\to \pi^0\nu\bar\nu$, we include angular correlation by the following Dalitz distribution, 
\begin{align}
\frac{d^2 \Gamma}{d q^2  d m_{\pi^0\nu}^2} \propto & f^2_+(q^2) \left[
m_{\pi^0\nu}^2(m_{K_L}^2+m_{\pi^0}^2-q^2 - m_{\pi^0\nu}^2) \right.\nonumber \\
& \left. \quad \quad \quad - m_{K_L}^2 m_{\pi^0}^2\right]\,,
\end{align}
where $q^2=m_{\nu\bar\nu}^2$, $m_{ij}^2 = (p_i + p_j)^2$, and the form factor  is taken from Ref.~\cite{Carrasco:2016kpy} up to order of $q^6$. 
The position and energy of a detected photon on the ECAL were smeared based on the  ECAL parameters \cite{Sato:2015yqa}, 
\begin{align}
\frac{\sigma_E}{E}&=0.99\% \oplus \frac{1.74\%}{\sqrt{E/\GeV}}\,,\\
{\sigma_{\rm position}}&=\frac{\sigma_x\oplus \sigma_y}{\sqrt{2}}=2.50{\rm mm}\oplus \frac{4.40}{\sqrt{E/\GeV}}{\rm mm}\,.
\end{align}
These smearings, especially the one for energy, led to the smearing of reconstructed vertices and thus all reconstructed 4-momenta. 

Then we applied the kinematical cuts used in the previous KOTO analysis of 2015 data \cite{Ahn:2018mvc}. In our simple framework, the veto and shower shape cuts were not included. Then, we applied cuts for the reconstructed vertex and momentum $p_T^{\pi^0}$  defined in the new analysis \cite{KOTOslides}. We calculated the cut efficiencies for both $\pi^0\nu\bar\nu$ and $\pi^0 X$ ($m_X$=1, 10, 20,..., 290, 300\,MeV), and calculated the ratio of efficiencies, Eq.~\eqref{eq:KLKOTO2body}. 

For validation, we applied  the same scheme for the signal region of Ref.~\cite{Ahn:2018mvc} in the reconstructions of vertices and transverse momenta $p_T^{\pi^0}$, 
and Fig.~\ref{fig:shape} shows a comparison of our result with Fig.~4 of \cite{Ahn:2018mvc}. They agree very well. 

The remaining samples in the new signal region \cite{KOTOslides} have an event-by-event information on the distance between the $K_L$ decay vertex and the ECAL and on the energy (boost factor) of the $X$. We used it to calculate the weight of Eq.~\eqref{eq:K2piinv}.

\bibliography{KOTOref}

\begin{thebibliography}{48}%
\makeatletter
\providecommand \@ifxundefined [1]{%
 \@ifx{#1\undefined}
}%
\providecommand \@ifnum [1]{%
 \ifnum #1\expandafter \@firstoftwo
 \else \expandafter \@secondoftwo
 \fi
}%
\providecommand \@ifx [1]{%
 \ifx #1\expandafter \@firstoftwo
 \else \expandafter \@secondoftwo
 \fi
}%
\providecommand \natexlab [1]{#1}%
\providecommand \enquote  [1]{``#1''}%
\providecommand \bibnamefont  [1]{#1}%
\providecommand \bibfnamefont [1]{#1}%
\providecommand \citenamefont [1]{#1}%
\providecommand \href@noop [0]{\@secondoftwo}%
\providecommand \href [0]{\begingroup \@sanitize@url \@href}%
\providecommand \@href[1]{\@@startlink{#1}\@@href}%
\providecommand \@@href[1]{\endgroup#1\@@endlink}%
\providecommand \@sanitize@url [0]{\catcode `\\12\catcode `\$12\catcode
  `\&12\catcode `\#12\catcode `\^12\catcode `\_12\catcode `\%12\relax}%
\providecommand \@@startlink[1]{}%
\providecommand \@@endlink[0]{}%
\providecommand \url  [0]{\begingroup\@sanitize@url \@url }%
\providecommand \@url [1]{\endgroup\@href {#1}{\urlprefix }}%
\providecommand \urlprefix  [0]{URL }%
\providecommand \Eprint [0]{\href }%
\providecommand \doibase [0]{http://dx.doi.org/}%
\providecommand \selectlanguage [0]{\@gobble}%
\providecommand \bibinfo  [0]{\@secondoftwo}%
\providecommand \bibfield  [0]{\@secondoftwo}%
\providecommand \translation [1]{[#1]}%
\providecommand \BibitemOpen [0]{}%
\providecommand \bibitemStop [0]{}%
\providecommand \bibitemNoStop [0]{.\EOS\space}%
\providecommand \EOS [0]{\spacefactor3000\relax}%
\providecommand \BibitemShut  [1]{\csname bibitem#1\endcsname}%
\let\auto@bib@innerbib\@empty
\bibitem [{\citenamefont {Littenberg}(1989)}]{Littenberg:1989ix}%
  \BibitemOpen
  \bibfield  {author} {\bibinfo {author} {\bibfnamefont {L.~S.}\ \bibnamefont
  {Littenberg}},\ }\href {\doibase 10.1103/PhysRevD.39.3322} {\bibfield
  {journal} {\bibinfo  {journal} {Phys. Rev.}\ }\textbf {\bibinfo {volume}
  {D39}},\ \bibinfo {pages} {3322} (\bibinfo {year} {1989})}\BibitemShut
  {NoStop}%
\bibitem [{\citenamefont {Glashow}\ \emph {et~al.}(1970)\citenamefont
  {Glashow}, \citenamefont {Iliopoulos},\ and\ \citenamefont
  {Maiani}}]{Glashow:1970gm}%
  \BibitemOpen
  \bibfield  {author} {\bibinfo {author} {\bibfnamefont {S.~L.}\ \bibnamefont
  {Glashow}}, \bibinfo {author} {\bibfnamefont {J.}~\bibnamefont {Iliopoulos}},
  \ and\ \bibinfo {author} {\bibfnamefont {L.}~\bibnamefont {Maiani}},\
  }\bibfield  {booktitle} {\emph {\bibinfo {booktitle} {{Meeting of the Italian
  School of Physics and Weak Interactions Bologna, Italy, April 26-28,
  1984}}},\ }\href {\doibase 10.1103/PhysRevD.2.1285} {\bibfield  {journal}
  {\bibinfo  {journal} {Phys. Rev.}\ }\textbf {\bibinfo {volume} {D2}},\
  \bibinfo {pages} {1285} (\bibinfo {year} {1970})}\BibitemShut {NoStop}%
\bibitem [{\citenamefont {Buras}\ \emph {et~al.}(2006)\citenamefont {Buras},
  \citenamefont {Gorbahn}, \citenamefont {Haisch},\ and\ \citenamefont
  {Nierste}}]{Buras:2006gb}%
  \BibitemOpen
  \bibfield  {author} {\bibinfo {author} {\bibfnamefont {A.~J.}\ \bibnamefont
  {Buras}}, \bibinfo {author} {\bibfnamefont {M.}~\bibnamefont {Gorbahn}},
  \bibinfo {author} {\bibfnamefont {U.}~\bibnamefont {Haisch}}, \ and\ \bibinfo
  {author} {\bibfnamefont {U.}~\bibnamefont {Nierste}},\ }\href {\doibase
  10.1007/JHEP11(2012)167, 10.1088/1126-6708/2006/11/002} {\bibfield  {journal}
  {\bibinfo  {journal} {JHEP}\ }\textbf {\bibinfo {volume} {11}},\ \bibinfo
  {pages} {002} (\bibinfo {year} {2006})},\ \bibinfo {note} {[Erratum:
  JHEP11,167(2012)]},\ \Eprint {http://arxiv.org/abs/hep-ph/0603079}
  {arXiv:hep-ph/0603079 [hep-ph]} \BibitemShut {NoStop}%
\bibitem [{\citenamefont {Brod}\ \emph {et~al.}(2011)\citenamefont {Brod},
  \citenamefont {Gorbahn},\ and\ \citenamefont {Stamou}}]{Brod:2010hi}%
  \BibitemOpen
  \bibfield  {author} {\bibinfo {author} {\bibfnamefont {J.}~\bibnamefont
  {Brod}}, \bibinfo {author} {\bibfnamefont {M.}~\bibnamefont {Gorbahn}}, \
  and\ \bibinfo {author} {\bibfnamefont {E.}~\bibnamefont {Stamou}},\ }\href
  {\doibase 10.1103/PhysRevD.83.034030} {\bibfield  {journal} {\bibinfo
  {journal} {Phys. Rev.}\ }\textbf {\bibinfo {volume} {D83}},\ \bibinfo {pages}
  {034030} (\bibinfo {year} {2011})},\ \Eprint {http://arxiv.org/abs/1009.0947}
  {arXiv:1009.0947 [hep-ph]} \BibitemShut {NoStop}%
\bibitem [{\citenamefont {Buras}\ \emph {et~al.}(2015)\citenamefont {Buras},
  \citenamefont {Buttazzo}, \citenamefont {Girrbach-Noe},\ and\ \citenamefont
  {Knegjens}}]{Buras:2015qea}%
  \BibitemOpen
  \bibfield  {author} {\bibinfo {author} {\bibfnamefont {A.~J.}\ \bibnamefont
  {Buras}}, \bibinfo {author} {\bibfnamefont {D.}~\bibnamefont {Buttazzo}},
  \bibinfo {author} {\bibfnamefont {J.}~\bibnamefont {Girrbach-Noe}}, \ and\
  \bibinfo {author} {\bibfnamefont {R.}~\bibnamefont {Knegjens}},\ }\href
  {\doibase 10.1007/JHEP11(2015)033} {\bibfield  {journal} {\bibinfo  {journal}
  {JHEP}\ }\textbf {\bibinfo {volume} {11}},\ \bibinfo {pages} {033} (\bibinfo
  {year} {2015})},\ \Eprint {http://arxiv.org/abs/1503.02693} {arXiv:1503.02693
  [hep-ph]} \BibitemShut {NoStop}%
\bibitem [{\citenamefont {Shinohara}()}]{KOTOslides}%
  \BibitemOpen
  \bibfield  {author} {\bibinfo {author} {\bibfnamefont {S.}~\bibnamefont
  {Shinohara}},\ }\href
  {https://indico.cern.ch/event/769729/contributions/3510939/attachments/1904988/3145907/KAON2019_shinohara_upload.pdf}
  {\enquote {\bibinfo {title} {{Search for the rare decay $K_L \to \pi^0 \nu
  \bar\nu $ at J-PARC KOTO experiment}},}\ }\bibinfo {note} {{KAON2019,
  Perugia, Italy, 10-13 September 2019}}\BibitemShut {NoStop}%
\bibitem [{\citenamefont {Ruggiero}()}]{NA62slides}%
  \BibitemOpen
  \bibfield  {author} {\bibinfo {author} {\bibfnamefont {G.}~\bibnamefont
  {Ruggiero}},\ }\href
  {https://indico.cern.ch/event/769729/contributions/3510938/attachments/1905346/3146619/kaon2019_ruggiero_final.pdf}
  {\enquote {\bibinfo {title} {{New Result on $K^+ \to \pi^+ \nu \bar\nu $ from
  the NA62 Experiment}},}\ }\bibinfo {note} {{KAON2019, Perugia, Italy, 10-13
  September 2019}}\BibitemShut {NoStop}%
\bibitem [{\citenamefont {Artamonov}\ \emph {et~al.}(2008)\citenamefont
  {Artamonov} \emph {et~al.}}]{Artamonov:2008qb}%
  \BibitemOpen
  \bibfield  {author} {\bibinfo {author} {\bibfnamefont {A.~V.}\ \bibnamefont
  {Artamonov}} \emph {et~al.} (\bibinfo {collaboration} {E949}),\ }\href
  {\doibase 10.1103/PhysRevLett.101.191802} {\bibfield  {journal} {\bibinfo
  {journal} {Phys. Rev. Lett.}\ }\textbf {\bibinfo {volume} {101}},\ \bibinfo
  {pages} {191802} (\bibinfo {year} {2008})},\ \Eprint
  {http://arxiv.org/abs/0808.2459} {arXiv:0808.2459 [hep-ex]} \BibitemShut
  {NoStop}%
\bibitem [{\citenamefont {Artamonov}\ \emph {et~al.}(2009)\citenamefont
  {Artamonov} \emph {et~al.}}]{Artamonov:2009sz}%
  \BibitemOpen
  \bibfield  {author} {\bibinfo {author} {\bibfnamefont {A.~V.}\ \bibnamefont
  {Artamonov}} \emph {et~al.} (\bibinfo {collaboration} {BNL-E949}),\ }\href
  {\doibase 10.1103/PhysRevD.79.092004} {\bibfield  {journal} {\bibinfo
  {journal} {Phys. Rev.}\ }\textbf {\bibinfo {volume} {D79}},\ \bibinfo {pages}
  {092004} (\bibinfo {year} {2009})},\ \Eprint {http://arxiv.org/abs/0903.0030}
  {arXiv:0903.0030 [hep-ex]} \BibitemShut {NoStop}%
\bibitem [{\citenamefont {Peccei}\ and\ \citenamefont
  {Quinn}(1977{\natexlab{a}})}]{Peccei:1977ur}%
  \BibitemOpen
  \bibfield  {author} {\bibinfo {author} {\bibfnamefont {R.~D.}\ \bibnamefont
  {Peccei}}\ and\ \bibinfo {author} {\bibfnamefont {H.~R.}\ \bibnamefont
  {Quinn}},\ }\href {\doibase 10.1103/PhysRevD.16.1791} {\bibfield  {journal}
  {\bibinfo  {journal} {Phys. Rev.}\ }\textbf {\bibinfo {volume} {D16}},\
  \bibinfo {pages} {1791} (\bibinfo {year} {1977}{\natexlab{a}})}\BibitemShut
  {NoStop}%
\bibitem [{\citenamefont {Peccei}\ and\ \citenamefont
  {Quinn}(1977{\natexlab{b}})}]{Peccei:1977hh}%
  \BibitemOpen
  \bibfield  {author} {\bibinfo {author} {\bibfnamefont {R.~D.}\ \bibnamefont
  {Peccei}}\ and\ \bibinfo {author} {\bibfnamefont {H.~R.}\ \bibnamefont
  {Quinn}},\ }\href {\doibase 10.1103/PhysRevLett.38.1440} {\bibfield
  {journal} {\bibinfo  {journal} {Phys. Rev. Lett.}\ }\textbf {\bibinfo
  {volume} {38}},\ \bibinfo {pages} {1440} (\bibinfo {year}
  {1977}{\natexlab{b}})},\ \bibinfo {note} {[,328(1977)]}\BibitemShut {NoStop}%
\bibitem [{\citenamefont {Weinberg}(1978)}]{Weinberg:1977ma}%
  \BibitemOpen
  \bibfield  {author} {\bibinfo {author} {\bibfnamefont {S.}~\bibnamefont
  {Weinberg}},\ }\href {\doibase 10.1103/PhysRevLett.40.223} {\bibfield
  {journal} {\bibinfo  {journal} {Phys. Rev. Lett.}\ }\textbf {\bibinfo
  {volume} {40}},\ \bibinfo {pages} {223} (\bibinfo {year} {1978})}\BibitemShut
  {NoStop}%
\bibitem [{\citenamefont {Wilczek}(1978)}]{Wilczek:1977pj}%
  \BibitemOpen
  \bibfield  {author} {\bibinfo {author} {\bibfnamefont {F.}~\bibnamefont
  {Wilczek}},\ }\href {\doibase 10.1103/PhysRevLett.40.279} {\bibfield
  {journal} {\bibinfo  {journal} {Phys. Rev. Lett.}\ }\textbf {\bibinfo
  {volume} {40}},\ \bibinfo {pages} {279} (\bibinfo {year} {1978})}\BibitemShut
  {NoStop}%
\bibitem [{\citenamefont {Graham}\ \emph {et~al.}(2015)\citenamefont {Graham},
  \citenamefont {Kaplan},\ and\ \citenamefont {Rajendran}}]{Graham:2015cka}%
  \BibitemOpen
  \bibfield  {author} {\bibinfo {author} {\bibfnamefont {P.~W.}\ \bibnamefont
  {Graham}}, \bibinfo {author} {\bibfnamefont {D.~E.}\ \bibnamefont {Kaplan}},
  \ and\ \bibinfo {author} {\bibfnamefont {S.}~\bibnamefont {Rajendran}},\
  }\href {\doibase 10.1103/PhysRevLett.115.221801} {\bibfield  {journal}
  {\bibinfo  {journal} {Phys. Rev. Lett.}\ }\textbf {\bibinfo {volume} {115}},\
  \bibinfo {pages} {221801} (\bibinfo {year} {2015})},\ \Eprint
  {http://arxiv.org/abs/1504.07551} {arXiv:1504.07551 [hep-ph]} \BibitemShut
  {NoStop}%
\bibitem [{\citenamefont {Frugiuele}\ \emph {et~al.}(2018)\citenamefont
  {Frugiuele}, \citenamefont {Fuchs}, \citenamefont {Perez},\ and\
  \citenamefont {Schlaffer}}]{Frugiuele:2018coc}%
  \BibitemOpen
  \bibfield  {author} {\bibinfo {author} {\bibfnamefont {C.}~\bibnamefont
  {Frugiuele}}, \bibinfo {author} {\bibfnamefont {E.}~\bibnamefont {Fuchs}},
  \bibinfo {author} {\bibfnamefont {G.}~\bibnamefont {Perez}}, \ and\ \bibinfo
  {author} {\bibfnamefont {M.}~\bibnamefont {Schlaffer}},\ }\href {\doibase
  10.1007/JHEP10(2018)151} {\bibfield  {journal} {\bibinfo  {journal} {JHEP}\
  }\textbf {\bibinfo {volume} {10}},\ \bibinfo {pages} {151} (\bibinfo {year}
  {2018})},\ \Eprint {http://arxiv.org/abs/1807.10842} {arXiv:1807.10842
  [hep-ph]} \BibitemShut {NoStop}%
\bibitem [{\citenamefont {Flacke}\ \emph {et~al.}(2017)\citenamefont {Flacke},
  \citenamefont {Frugiuele}, \citenamefont {Fuchs}, \citenamefont {Gupta},\
  and\ \citenamefont {Perez}}]{Flacke:2016szy}%
  \BibitemOpen
  \bibfield  {author} {\bibinfo {author} {\bibfnamefont {T.}~\bibnamefont
  {Flacke}}, \bibinfo {author} {\bibfnamefont {C.}~\bibnamefont {Frugiuele}},
  \bibinfo {author} {\bibfnamefont {E.}~\bibnamefont {Fuchs}}, \bibinfo
  {author} {\bibfnamefont {R.~S.}\ \bibnamefont {Gupta}}, \ and\ \bibinfo
  {author} {\bibfnamefont {G.}~\bibnamefont {Perez}},\ }\href {\doibase
  10.1007/JHEP06(2017)050} {\bibfield  {journal} {\bibinfo  {journal} {JHEP}\
  }\textbf {\bibinfo {volume} {06}},\ \bibinfo {pages} {050} (\bibinfo {year}
  {2017})},\ \Eprint {http://arxiv.org/abs/1610.02025} {arXiv:1610.02025
  [hep-ph]} \BibitemShut {NoStop}%
\bibitem [{\citenamefont {Grossman}\ and\ \citenamefont
  {Nir}(1997)}]{Grossman:1997sk}%
  \BibitemOpen
  \bibfield  {author} {\bibinfo {author} {\bibfnamefont {Y.}~\bibnamefont
  {Grossman}}\ and\ \bibinfo {author} {\bibfnamefont {Y.}~\bibnamefont {Nir}},\
  }\href {\doibase 10.1016/S0370-2693(97)00210-4} {\bibfield  {journal}
  {\bibinfo  {journal} {Phys. Lett.}\ }\textbf {\bibinfo {volume} {B398}},\
  \bibinfo {pages} {163} (\bibinfo {year} {1997})},\ \Eprint
  {http://arxiv.org/abs/hep-ph/9701313} {arXiv:hep-ph/9701313 [hep-ph]}
  \BibitemShut {NoStop}%
\bibitem [{\citenamefont {Mescia}\ and\ \citenamefont
  {Smith}(2007)}]{Mescia:2007kn}%
  \BibitemOpen
  \bibfield  {author} {\bibinfo {author} {\bibfnamefont {F.}~\bibnamefont
  {Mescia}}\ and\ \bibinfo {author} {\bibfnamefont {C.}~\bibnamefont {Smith}},\
  }\href {\doibase 10.1103/PhysRevD.76.034017} {\bibfield  {journal} {\bibinfo
  {journal} {Phys. Rev.}\ }\textbf {\bibinfo {volume} {D76}},\ \bibinfo {pages}
  {034017} (\bibinfo {year} {2007})},\ \Eprint {http://arxiv.org/abs/0705.2025}
  {arXiv:0705.2025 [hep-ph]} \BibitemShut {NoStop}%
\bibitem [{\citenamefont {Tanabashi}\ \emph {et~al.}(2018)\citenamefont
  {Tanabashi} \emph {et~al.}}]{Tanabashi:2018oca}%
  \BibitemOpen
  \bibfield  {author} {\bibinfo {author} {\bibfnamefont {M.}~\bibnamefont
  {Tanabashi}} \emph {et~al.} (\bibinfo {collaboration} {Particle Data
  Group}),\ }\href {\doibase 10.1103/PhysRevD.98.030001} {\bibfield  {journal}
  {\bibinfo  {journal} {Phys. Rev.}\ }\textbf {\bibinfo {volume} {D98}},\
  \bibinfo {pages} {030001} (\bibinfo {year} {2018})}\BibitemShut {NoStop}%
\bibitem [{\citenamefont {Abouzaid}\ \emph {et~al.}(2008)\citenamefont
  {Abouzaid} \emph {et~al.}}]{Abouzaid:2008xm}%
  \BibitemOpen
  \bibfield  {author} {\bibinfo {author} {\bibfnamefont {E.}~\bibnamefont
  {Abouzaid}} \emph {et~al.} (\bibinfo {collaboration} {KTeV}),\ }\href
  {\doibase 10.1103/PhysRevD.77.112004} {\bibfield  {journal} {\bibinfo
  {journal} {Phys. Rev.}\ }\textbf {\bibinfo {volume} {D77}},\ \bibinfo {pages}
  {112004} (\bibinfo {year} {2008})},\ \Eprint {http://arxiv.org/abs/0805.0031}
  {arXiv:0805.0031 [hep-ex]} \BibitemShut {NoStop}%
\bibitem [{\citenamefont {Alavi-Harati}\ \emph {et~al.}(2004)\citenamefont
  {Alavi-Harati} \emph {et~al.}}]{AlaviHarati:2003mr}%
  \BibitemOpen
  \bibfield  {author} {\bibinfo {author} {\bibfnamefont {A.}~\bibnamefont
  {Alavi-Harati}} \emph {et~al.} (\bibinfo {collaboration} {KTeV}),\ }\href
  {\doibase 10.1103/PhysRevLett.93.021805} {\bibfield  {journal} {\bibinfo
  {journal} {Phys. Rev. Lett.}\ }\textbf {\bibinfo {volume} {93}},\ \bibinfo
  {pages} {021805} (\bibinfo {year} {2004})},\ \Eprint
  {http://arxiv.org/abs/hep-ex/0309072} {arXiv:hep-ex/0309072 [hep-ex]}
  \BibitemShut {NoStop}%
\bibitem [{\citenamefont {Alavi-Harati}\ \emph {et~al.}(2000)\citenamefont
  {Alavi-Harati} \emph {et~al.}}]{AlaviHarati:2000hs}%
  \BibitemOpen
  \bibfield  {author} {\bibinfo {author} {\bibfnamefont {A.}~\bibnamefont
  {Alavi-Harati}} \emph {et~al.} (\bibinfo {collaboration} {KTEV}),\ }\href
  {\doibase 10.1103/PhysRevLett.84.5279} {\bibfield  {journal} {\bibinfo
  {journal} {Phys. Rev. Lett.}\ }\textbf {\bibinfo {volume} {84}},\ \bibinfo
  {pages} {5279} (\bibinfo {year} {2000})},\ \Eprint
  {http://arxiv.org/abs/hep-ex/0001006} {arXiv:hep-ex/0001006 [hep-ex]}
  \BibitemShut {NoStop}%
\bibitem [{\citenamefont {Pernas}()}]{LHCbslides}%
  \BibitemOpen
  \bibfield  {author} {\bibinfo {author} {\bibfnamefont {M.~R.}\ \bibnamefont
  {Pernas}},\ }\href
  {https://indico.cern.ch/event/769729/contributions/3510936/attachments/1904808/3154425/Miguel_Ramos_Pernas_K0S2mu2.pdf}
  {\enquote {\bibinfo {title} {{Search for $K_S^0 \to \mu^+ \mu^-$ at LHCb}},}\
  }\bibinfo {note} {{KAON2019, Perugia, Italy, 10-13 September
  2019}}\BibitemShut {NoStop}%
\bibitem [{\citenamefont {D'Ambrosio}\ and\ \citenamefont
  {Kitahara}(2017)}]{DAmbrosio:2017klp}%
  \BibitemOpen
  \bibfield  {author} {\bibinfo {author} {\bibfnamefont {G.}~\bibnamefont
  {D'Ambrosio}}\ and\ \bibinfo {author} {\bibfnamefont {T.}~\bibnamefont
  {Kitahara}},\ }\href {\doibase 10.1103/PhysRevLett.119.201802} {\bibfield
  {journal} {\bibinfo  {journal} {Phys. Rev. Lett.}\ }\textbf {\bibinfo
  {volume} {119}},\ \bibinfo {pages} {201802} (\bibinfo {year} {2017})},\
  \Eprint {http://arxiv.org/abs/1707.06999} {arXiv:1707.06999 [hep-ph]}
  \BibitemShut {NoStop}%
\bibitem [{\citenamefont {Chobanova}\ \emph {et~al.}(2018)\citenamefont
  {Chobanova}, \citenamefont {D'Ambrosio}, \citenamefont {Kitahara},
  \citenamefont {Lucio~Martinez}, \citenamefont {Martinez~Santos},
  \citenamefont {Fernandez},\ and\ \citenamefont
  {Yamamoto}}]{Chobanova:2017rkj}%
  \BibitemOpen
  \bibfield  {author} {\bibinfo {author} {\bibfnamefont {V.}~\bibnamefont
  {Chobanova}}, \bibinfo {author} {\bibfnamefont {G.}~\bibnamefont
  {D'Ambrosio}}, \bibinfo {author} {\bibfnamefont {T.}~\bibnamefont
  {Kitahara}}, \bibinfo {author} {\bibfnamefont {M.}~\bibnamefont
  {Lucio~Martinez}}, \bibinfo {author} {\bibfnamefont {D.}~\bibnamefont
  {Martinez~Santos}}, \bibinfo {author} {\bibfnamefont {I.~S.}\ \bibnamefont
  {Fernandez}}, \ and\ \bibinfo {author} {\bibfnamefont {K.}~\bibnamefont
  {Yamamoto}},\ }\href {\doibase 10.1007/JHEP05(2018)024} {\bibfield  {journal}
  {\bibinfo  {journal} {JHEP}\ }\textbf {\bibinfo {volume} {05}},\ \bibinfo
  {pages} {024} (\bibinfo {year} {2018})},\ \Eprint
  {http://arxiv.org/abs/1711.11030} {arXiv:1711.11030 [hep-ph]} \BibitemShut
  {NoStop}%
\bibitem [{\citenamefont {Agashe}\ \emph {et~al.}(2006)\citenamefont {Agashe},
  \citenamefont {Contino}, \citenamefont {Da~Rold},\ and\ \citenamefont
  {Pomarol}}]{Agashe:2006at}%
  \BibitemOpen
  \bibfield  {author} {\bibinfo {author} {\bibfnamefont {K.}~\bibnamefont
  {Agashe}}, \bibinfo {author} {\bibfnamefont {R.}~\bibnamefont {Contino}},
  \bibinfo {author} {\bibfnamefont {L.}~\bibnamefont {Da~Rold}}, \ and\
  \bibinfo {author} {\bibfnamefont {A.}~\bibnamefont {Pomarol}},\ }\href
  {\doibase 10.1016/j.physletb.2006.08.005} {\bibfield  {journal} {\bibinfo
  {journal} {Phys. Lett.}\ }\textbf {\bibinfo {volume} {B641}},\ \bibinfo
  {pages} {62} (\bibinfo {year} {2006})},\ \Eprint
  {http://arxiv.org/abs/hep-ph/0605341} {arXiv:hep-ph/0605341 [hep-ph]}
  \BibitemShut {NoStop}%
\bibitem [{\citenamefont {Grossman}\ and\ \citenamefont
  {Nir}(2012)}]{Grossman:2011zk}%
  \BibitemOpen
  \bibfield  {author} {\bibinfo {author} {\bibfnamefont {Y.}~\bibnamefont
  {Grossman}}\ and\ \bibinfo {author} {\bibfnamefont {Y.}~\bibnamefont {Nir}},\
  }\href {\doibase 10.1007/JHEP04(2012)002} {\bibfield  {journal} {\bibinfo
  {journal} {JHEP}\ }\textbf {\bibinfo {volume} {04}},\ \bibinfo {pages} {002}
  (\bibinfo {year} {2012})},\ \Eprint {http://arxiv.org/abs/1110.3790}
  {arXiv:1110.3790 [hep-ph]} \BibitemShut {NoStop}%
\bibitem [{\citenamefont {Cirigliano}\ \emph {et~al.}(2018)\citenamefont
  {Cirigliano}, \citenamefont {Crivellin},\ and\ \citenamefont
  {Hoferichter}}]{Cirigliano:2017tqn}%
  \BibitemOpen
  \bibfield  {author} {\bibinfo {author} {\bibfnamefont {V.}~\bibnamefont
  {Cirigliano}}, \bibinfo {author} {\bibfnamefont {A.}~\bibnamefont
  {Crivellin}}, \ and\ \bibinfo {author} {\bibfnamefont {M.}~\bibnamefont
  {Hoferichter}},\ }\href {\doibase 10.1103/PhysRevLett.120.141803} {\bibfield
  {journal} {\bibinfo  {journal} {Phys. Rev. Lett.}\ }\textbf {\bibinfo
  {volume} {120}},\ \bibinfo {pages} {141803} (\bibinfo {year} {2018})},\
  \Eprint {http://arxiv.org/abs/1712.06595} {arXiv:1712.06595 [hep-ph]}
  \BibitemShut {NoStop}%
\bibitem [{\citenamefont {Gedalia}\ \emph {et~al.}(2012)\citenamefont
  {Gedalia}, \citenamefont {Kamenik}, \citenamefont {Ligeti},\ and\
  \citenamefont {Perez}}]{Gedalia:2012pi}%
  \BibitemOpen
  \bibfield  {author} {\bibinfo {author} {\bibfnamefont {O.}~\bibnamefont
  {Gedalia}}, \bibinfo {author} {\bibfnamefont {J.~F.}\ \bibnamefont
  {Kamenik}}, \bibinfo {author} {\bibfnamefont {Z.}~\bibnamefont {Ligeti}}, \
  and\ \bibinfo {author} {\bibfnamefont {G.}~\bibnamefont {Perez}},\ }\href
  {\doibase 10.1016/j.physletb.2012.06.050} {\bibfield  {journal} {\bibinfo
  {journal} {Phys. Lett.}\ }\textbf {\bibinfo {volume} {B714}},\ \bibinfo
  {pages} {55} (\bibinfo {year} {2012})},\ \Eprint
  {http://arxiv.org/abs/1202.5038} {arXiv:1202.5038 [hep-ph]} \BibitemShut
  {NoStop}%
\bibitem [{\citenamefont {Ahn}\ \emph {et~al.}(2019)\citenamefont {Ahn} \emph
  {et~al.}}]{Ahn:2018mvc}%
  \BibitemOpen
  \bibfield  {author} {\bibinfo {author} {\bibfnamefont {J.~K.}\ \bibnamefont
  {Ahn}} \emph {et~al.} (\bibinfo {collaboration} {KOTO}),\ }\href {\doibase
  10.1103/PhysRevLett.122.021802} {\bibfield  {journal} {\bibinfo  {journal}
  {Phys. Rev. Lett.}\ }\textbf {\bibinfo {volume} {122}},\ \bibinfo {pages}
  {021802} (\bibinfo {year} {2019})},\ \Eprint
  {http://arxiv.org/abs/1810.09655} {arXiv:1810.09655 [hep-ex]} \BibitemShut
  {NoStop}%
\bibitem [{\citenamefont {Leutwyler}\ and\ \citenamefont
  {Shifman}(1990)}]{Leutwyler:1989xj}%
  \BibitemOpen
  \bibfield  {author} {\bibinfo {author} {\bibfnamefont {H.}~\bibnamefont
  {Leutwyler}}\ and\ \bibinfo {author} {\bibfnamefont {M.~A.}\ \bibnamefont
  {Shifman}},\ }\href {\doibase 10.1016/0550-3213(90)90475-S} {\bibfield
  {journal} {\bibinfo  {journal} {Nucl. Phys.}\ }\textbf {\bibinfo {volume}
  {B343}},\ \bibinfo {pages} {369} (\bibinfo {year} {1990})}\BibitemShut
  {NoStop}%
\bibitem [{\citenamefont {Cortina~Gil}\ \emph {et~al.}(2019)\citenamefont
  {Cortina~Gil} \emph {et~al.}}]{CortinaGil:2018fkc}%
  \BibitemOpen
  \bibfield  {author} {\bibinfo {author} {\bibfnamefont {E.}~\bibnamefont
  {Cortina~Gil}} \emph {et~al.} (\bibinfo {collaboration} {NA62}),\ }\href
  {\doibase 10.1016/j.physletb.2019.01.067} {\bibfield  {journal} {\bibinfo
  {journal} {Phys. Lett.}\ }\textbf {\bibinfo {volume} {B791}},\ \bibinfo
  {pages} {156} (\bibinfo {year} {2019})},\ \Eprint
  {http://arxiv.org/abs/1811.08508} {arXiv:1811.08508 [hep-ex]} \BibitemShut
  {NoStop}%
\bibitem [{\citenamefont {Fuyuto}\ \emph {et~al.}(2015)\citenamefont {Fuyuto},
  \citenamefont {Hou},\ and\ \citenamefont {Kohda}}]{Fuyuto:2014cya}%
  \BibitemOpen
  \bibfield  {author} {\bibinfo {author} {\bibfnamefont {K.}~\bibnamefont
  {Fuyuto}}, \bibinfo {author} {\bibfnamefont {W.-S.}\ \bibnamefont {Hou}}, \
  and\ \bibinfo {author} {\bibfnamefont {M.}~\bibnamefont {Kohda}},\ }\href
  {\doibase 10.1103/PhysRevLett.114.171802} {\bibfield  {journal} {\bibinfo
  {journal} {Phys. Rev. Lett.}\ }\textbf {\bibinfo {volume} {114}},\ \bibinfo
  {pages} {171802} (\bibinfo {year} {2015})},\ \Eprint
  {http://arxiv.org/abs/1412.4397} {arXiv:1412.4397 [hep-ph]} \BibitemShut
  {NoStop}%
\bibitem [{\citenamefont {Artamonov}\ \emph {et~al.}(2005)\citenamefont
  {Artamonov} \emph {et~al.}}]{Artamonov:2005ru}%
  \BibitemOpen
  \bibfield  {author} {\bibinfo {author} {\bibfnamefont {A.~V.}\ \bibnamefont
  {Artamonov}} \emph {et~al.} (\bibinfo {collaboration} {E949}),\ }\href
  {\doibase 10.1016/j.physletb.2005.07.057} {\bibfield  {journal} {\bibinfo
  {journal} {Phys. Lett.}\ }\textbf {\bibinfo {volume} {B623}},\ \bibinfo
  {pages} {192} (\bibinfo {year} {2005})},\ \Eprint
  {http://arxiv.org/abs/hep-ex/0505069} {arXiv:hep-ex/0505069 [hep-ex]}
  \BibitemShut {NoStop}%
\bibitem [{\citenamefont {Lazzeroni}\ \emph {et~al.}(2014)\citenamefont
  {Lazzeroni} \emph {et~al.}}]{Ceccucci:2014oza}%
  \BibitemOpen
  \bibfield  {author} {\bibinfo {author} {\bibfnamefont {C.}~\bibnamefont
  {Lazzeroni}} \emph {et~al.} (\bibinfo {collaboration} {NA62}),\ }\href
  {\doibase 10.1016/j.physletb.2014.03.016} {\bibfield  {journal} {\bibinfo
  {journal} {Phys. Lett.}\ }\textbf {\bibinfo {volume} {B732}},\ \bibinfo
  {pages} {65} (\bibinfo {year} {2014})},\ \Eprint
  {http://arxiv.org/abs/1402.4334} {arXiv:1402.4334 [hep-ex]} \BibitemShut
  {NoStop}%
\bibitem [{\citenamefont {Gori}\ \emph {et~al.}()\citenamefont {Gori},
  \citenamefont {Perez},\ and\ \citenamefont {Tobioka}}]{GORIPEREZTOBIOKA}%
  \BibitemOpen
  \bibfield  {author} {\bibinfo {author} {\bibfnamefont {S.}~\bibnamefont
  {Gori}}, \bibinfo {author} {\bibfnamefont {G.}~\bibnamefont {Perez}}, \ and\
  \bibinfo {author} {\bibfnamefont {K.}~\bibnamefont {Tobioka}},\ }\href@noop
  {} {}\bibinfo {note} {{in preparation}}\BibitemShut {NoStop}%
\bibitem [{\citenamefont {Beacham}\ \emph {et~al.}(2019)\citenamefont {Beacham}
  \emph {et~al.}}]{Beacham:2019nyx}%
  \BibitemOpen
  \bibfield  {author} {\bibinfo {author} {\bibfnamefont {J.}~\bibnamefont
  {Beacham}} \emph {et~al.},\ }\href@noop {} {\  (\bibinfo {year} {2019})},\
  \Eprint {http://arxiv.org/abs/1901.09966} {arXiv:1901.09966 [hep-ex]}
  \BibitemShut {NoStop}%
\bibitem [{\citenamefont {Georgi}\ \emph {et~al.}(1986)\citenamefont {Georgi},
  \citenamefont {Kaplan},\ and\ \citenamefont {Randall}}]{Georgi:1986df}%
  \BibitemOpen
  \bibfield  {author} {\bibinfo {author} {\bibfnamefont {H.}~\bibnamefont
  {Georgi}}, \bibinfo {author} {\bibfnamefont {D.~B.}\ \bibnamefont {Kaplan}},
  \ and\ \bibinfo {author} {\bibfnamefont {L.}~\bibnamefont {Randall}},\ }\href
  {\doibase 10.1016/0370-2693(86)90688-X} {\bibfield  {journal} {\bibinfo
  {journal} {Phys. Lett.}\ }\textbf {\bibinfo {volume} {169B}},\ \bibinfo
  {pages} {73} (\bibinfo {year} {1986})}\BibitemShut {NoStop}%
\bibitem [{\citenamefont {Bauer}\ \emph {et~al.}(2017)\citenamefont {Bauer},
  \citenamefont {Neubert},\ and\ \citenamefont {Thamm}}]{Bauer:2017ris}%
  \BibitemOpen
  \bibfield  {author} {\bibinfo {author} {\bibfnamefont {M.}~\bibnamefont
  {Bauer}}, \bibinfo {author} {\bibfnamefont {M.}~\bibnamefont {Neubert}}, \
  and\ \bibinfo {author} {\bibfnamefont {A.}~\bibnamefont {Thamm}},\ }\href
  {\doibase 10.1007/JHEP12(2017)044} {\bibfield  {journal} {\bibinfo  {journal}
  {JHEP}\ }\textbf {\bibinfo {volume} {12}},\ \bibinfo {pages} {044} (\bibinfo
  {year} {2017})},\ \Eprint {http://arxiv.org/abs/1708.00443} {arXiv:1708.00443
  [hep-ph]} \BibitemShut {NoStop}%
\bibitem [{\citenamefont {Aloni}\ \emph
  {et~al.}(2019{\natexlab{a}})\citenamefont {Aloni}, \citenamefont {Soreq},\
  and\ \citenamefont {Williams}}]{Aloni:2018vki}%
  \BibitemOpen
  \bibfield  {author} {\bibinfo {author} {\bibfnamefont {D.}~\bibnamefont
  {Aloni}}, \bibinfo {author} {\bibfnamefont {Y.}~\bibnamefont {Soreq}}, \ and\
  \bibinfo {author} {\bibfnamefont {M.}~\bibnamefont {Williams}},\ }\href
  {\doibase 10.1103/PhysRevLett.123.031803} {\bibfield  {journal} {\bibinfo
  {journal} {Phys. Rev. Lett.}\ }\textbf {\bibinfo {volume} {123}},\ \bibinfo
  {pages} {031803} (\bibinfo {year} {2019}{\natexlab{a}})},\ \Eprint
  {http://arxiv.org/abs/1811.03474} {arXiv:1811.03474 [hep-ph]} \BibitemShut
  {NoStop}%
\bibitem [{\citenamefont {Masuda}\ \emph {et~al.}(2016)\citenamefont {Masuda}
  \emph {et~al.}}]{Masuda:2015eta}%
  \BibitemOpen
  \bibfield  {author} {\bibinfo {author} {\bibfnamefont {T.}~\bibnamefont
  {Masuda}} \emph {et~al.},\ }\href {\doibase 10.1093/ptep/ptv171} {\bibfield
  {journal} {\bibinfo  {journal} {PTEP}\ }\textbf {\bibinfo {volume} {2016}},\
  \bibinfo {pages} {013C03} (\bibinfo {year} {2016})},\ \Eprint
  {http://arxiv.org/abs/1509.03386} {arXiv:1509.03386 [physics.ins-det]}
  \BibitemShut {NoStop}%
\bibitem [{\citenamefont {Aloni}\ \emph
  {et~al.}(2019{\natexlab{b}})\citenamefont {Aloni}, \citenamefont {Fanelli},
  \citenamefont {Soreq},\ and\ \citenamefont {Williams}}]{Aloni:2019ruo}%
  \BibitemOpen
  \bibfield  {author} {\bibinfo {author} {\bibfnamefont {D.}~\bibnamefont
  {Aloni}}, \bibinfo {author} {\bibfnamefont {C.}~\bibnamefont {Fanelli}},
  \bibinfo {author} {\bibfnamefont {Y.}~\bibnamefont {Soreq}}, \ and\ \bibinfo
  {author} {\bibfnamefont {M.}~\bibnamefont {Williams}},\ }\href {\doibase
  10.1103/PhysRevLett.123.071801} {\bibfield  {journal} {\bibinfo  {journal}
  {Phys. Rev. Lett.}\ }\textbf {\bibinfo {volume} {123}},\ \bibinfo {pages}
  {071801} (\bibinfo {year} {2019}{\natexlab{b}})},\ \Eprint
  {http://arxiv.org/abs/1903.03586} {arXiv:1903.03586 [hep-ph]} \BibitemShut
  {NoStop}%
\bibitem [{\citenamefont {Ebadi}\ \emph {et~al.}(2019)\citenamefont {Ebadi},
  \citenamefont {Khatibi},\ and\ \citenamefont
  {Mohammadi~Najafabadi}}]{Ebadi:2019gij}%
  \BibitemOpen
  \bibfield  {author} {\bibinfo {author} {\bibfnamefont {J.}~\bibnamefont
  {Ebadi}}, \bibinfo {author} {\bibfnamefont {S.}~\bibnamefont {Khatibi}}, \
  and\ \bibinfo {author} {\bibfnamefont {M.}~\bibnamefont
  {Mohammadi~Najafabadi}},\ }\href {\doibase 10.1103/PhysRevD.100.015016}
  {\bibfield  {journal} {\bibinfo  {journal} {Phys. Rev.}\ }\textbf {\bibinfo
  {volume} {D100}},\ \bibinfo {pages} {015016} (\bibinfo {year} {2019})},\
  \Eprint {http://arxiv.org/abs/1901.03061} {arXiv:1901.03061 [hep-ph]}
  \BibitemShut {NoStop}%
\bibitem [{\citenamefont {Altmannshofer}\ \emph {et~al.}(2019)\citenamefont
  {Altmannshofer}, \citenamefont {Gori},\ and\ \citenamefont
  {Robinson}}]{Altmannshofer:2019yji}%
  \BibitemOpen
  \bibfield  {author} {\bibinfo {author} {\bibfnamefont {W.}~\bibnamefont
  {Altmannshofer}}, \bibinfo {author} {\bibfnamefont {S.}~\bibnamefont {Gori}},
  \ and\ \bibinfo {author} {\bibfnamefont {D.~J.}\ \bibnamefont {Robinson}},\
  }\href@noop {} {\  (\bibinfo {year} {2019})},\ \Eprint
  {http://arxiv.org/abs/1909.00005} {arXiv:1909.00005 [hep-ph]} \BibitemShut
  {NoStop}%
\bibitem [{\citenamefont {Bergsma}\ \emph {et~al.}(1985)\citenamefont {Bergsma}
  \emph {et~al.}}]{Bergsma:1985qz}%
  \BibitemOpen
  \bibfield  {author} {\bibinfo {author} {\bibfnamefont {F.}~\bibnamefont
  {Bergsma}} \emph {et~al.} (\bibinfo {collaboration} {CHARM}),\ }\href
  {\doibase 10.1016/0370-2693(85)90400-9} {\bibfield  {journal} {\bibinfo
  {journal} {Phys. Lett.}\ }\textbf {\bibinfo {volume} {157B}},\ \bibinfo
  {pages} {458} (\bibinfo {year} {1985})}\BibitemShut {NoStop}%
\bibitem [{\citenamefont {Blumlein}\ \emph {et~al.}(1991)\citenamefont
  {Blumlein} \emph {et~al.}}]{Blumlein:1990ay}%
  \BibitemOpen
  \bibfield  {author} {\bibinfo {author} {\bibfnamefont {J.}~\bibnamefont
  {Blumlein}} \emph {et~al.},\ }\href {\doibase 10.1007/BF01548556} {\bibfield
  {journal} {\bibinfo  {journal} {Z. Phys.}\ }\textbf {\bibinfo {volume}
  {C51}},\ \bibinfo {pages} {341} (\bibinfo {year} {1991})}\BibitemShut
  {NoStop}%
\bibitem [{\citenamefont {Carrasco}\ \emph {et~al.}(2016)\citenamefont
  {Carrasco}, \citenamefont {Lami}, \citenamefont {Lubicz}, \citenamefont
  {Riggio}, \citenamefont {Simula},\ and\ \citenamefont
  {Tarantino}}]{Carrasco:2016kpy}%
  \BibitemOpen
  \bibfield  {author} {\bibinfo {author} {\bibfnamefont {N.}~\bibnamefont
  {Carrasco}}, \bibinfo {author} {\bibfnamefont {P.}~\bibnamefont {Lami}},
  \bibinfo {author} {\bibfnamefont {V.}~\bibnamefont {Lubicz}}, \bibinfo
  {author} {\bibfnamefont {L.}~\bibnamefont {Riggio}}, \bibinfo {author}
  {\bibfnamefont {S.}~\bibnamefont {Simula}}, \ and\ \bibinfo {author}
  {\bibfnamefont {C.}~\bibnamefont {Tarantino}},\ }\href {\doibase
  10.1103/PhysRevD.93.114512} {\bibfield  {journal} {\bibinfo  {journal} {Phys.
  Rev.}\ }\textbf {\bibinfo {volume} {D93}},\ \bibinfo {pages} {114512}
  (\bibinfo {year} {2016})},\ \Eprint {http://arxiv.org/abs/1602.04113}
  {arXiv:1602.04113 [hep-lat]} \BibitemShut {NoStop}%
\bibitem [{\citenamefont {Sato}\ \emph {et~al.}(2015)\citenamefont {Sato},
  \citenamefont {Iwai}, \citenamefont {Shiomi}, \citenamefont {Sugiyama},
  \citenamefont {Togawa},\ and\ \citenamefont {Yamanaka}}]{Sato:2015yqa}%
  \BibitemOpen
  \bibfield  {author} {\bibinfo {author} {\bibfnamefont {K.}~\bibnamefont
  {Sato}}, \bibinfo {author} {\bibfnamefont {E.}~\bibnamefont {Iwai}}, \bibinfo
  {author} {\bibfnamefont {K.}~\bibnamefont {Shiomi}}, \bibinfo {author}
  {\bibfnamefont {Y.}~\bibnamefont {Sugiyama}}, \bibinfo {author}
  {\bibfnamefont {M.}~\bibnamefont {Togawa}}, \ and\ \bibinfo {author}
  {\bibfnamefont {T.}~\bibnamefont {Yamanaka}},\ }\bibfield  {booktitle} {\emph
  {\bibinfo {booktitle} {{Proceedings, 2nd International Symposium on Science
  at J-PARC: Unlocking the Mysteries of Life, Matter and the Universe (J-PARC
  2014): Tsukuba, Japan, July 12-15, 2014}}},\ }\href {\doibase
  10.7566/JPSCP.8.024007} {\bibfield  {journal} {\bibinfo  {journal} {JPS Conf.
  Proc.}\ }\textbf {\bibinfo {volume} {8}},\ \bibinfo {pages} {024007}
  (\bibinfo {year} {2015})}\BibitemShut {NoStop}%
\end{thebibliography}%

\end{document}